\documentstyle[aps,pre]{revtex}

\begin{document}

\draft

\title{{\bf General theory of instabilities for patterns with sharp
    interfaces in reaction-diffusion systems}} 
\author{C. B. Muratov}
\address{Department of Physics, Boston University, Boston, MA 02215}
\author{V. V. Osipov} 
\address{Department of Theoretical Physics, Russian Science Center\\
  ``Orion''\\ 2/46 Plekhanov St., Moscow 111123, Russia} 
\date{\today} 
\maketitle
 
\begin{abstract} 
  An asymptotic method for finding instabilities of arbitrary
  $d$-dimensional large-amplitude patterns in a wide class of
  reaction-diffusion systems is presented. The complete stability
  analysis of 2- and 3-dimensional localized patterns is carried out.
  It is shown that in the considered class of systems the criteria for
  different types of instabilities are universal. The specific
  nonlinearities enter the criteria only via three numerical constants
  of order one. The performed analysis explains the self-organization
  scenarios observed in the recent experiments and numerical simulations
  of some concrete reaction-diffusion systems.
 
\end{abstract}


\section{introduction}

In the last two decades the problem of pattern formation and
self-organization has become a paradigm of modern science
\cite{nicolis,cross93,field,belintsev,ko:book,ko:ufn89,%
  ko:ufn90,ko:irreversible}. Patterns are observed in a wide variety of
physical systems, such as gas and electron-hole plasma; various
semiconductor, superconductor and gas-discharge structures; some
ferroelectric, magnetic, and optical media; systems with uniformly
generated combustion material (see
\cite{ko:book,ko:ufn89,ko:ufn90,bode95} and references therein). Pattern
formation and self-organization are most conspicuous in chemical and
biological systems (see \cite{nicolis,cross93,field,belintsev,ko:book}
and references therein). As a rule, all these systems are extremely
complicated. In order to describe pattern formation phenomena in them a
number of simplifications is made. The majority of the simplified models
reduce to a pair of reaction-diffusion equations of the
activator-inhibitor type \cite{ko:book,ko:ufn89,ko:ufn90}
\begin{equation} \label{1} \tau_\theta {\partial \theta \over \partial
    t} = l^2 \Delta \theta - q(\theta, \eta, A),
\end{equation} 
\begin{equation}
  \label{2} \tau_\eta {\partial \eta \over \partial t} = L^2 \Delta \eta 
  - Q(\theta, \eta, A),
\end{equation} 
where $\theta$ and $\eta$ are the distributions of the activator and the
inhibitor, respectively; $q(\theta, \eta, A)$ and $Q(\theta, \eta, A)$
are certain nonlinear functions; $l$ and $L$ are the characteristic
length scales, and $\tau_\theta$ and $\tau_\eta$ are the characteristic
time scales of $\theta$ and $\eta$, respectively; $A$ is the control
parameter. The well-known models of certain autocatalytic reactions,
such as the Brusselator \cite{nicolis}, the two-component version of the
Oregonator \cite{cross93} and the Gray-Scott \cite{gray} models, the
classical model of morphogenesis proposed by Gierer and Meinhardt
\cite{gierer}; the piecewise-linear \cite{rinzel73,koga80} and
FitzHugh-Nagumo \cite{fitz,nagumo} models describing the propagation of
impulses in the nerve tissue are the special cases of Eqs. (\ref{1}) and
(\ref{2}). These models are most widely used in the analytical
investigations of different types of patterns
\cite{gierer,rinzel73,koga80,fitz,nagumo,kko:mk84:1,kko:mk84:2,%
kko:dan84,ito92,ermentrout84,klaasen84,pearson:sci93,petrich94,%
krischer94,kuo:pre95,om:prl95}.

The main self-organization phenomenon in the considered systems is
spontaneous transformation of one type of pattern to the other as
certain parameters of the system are varied. Self-organization scenarios
become extremely diverse in 2- and 3-dimensional systems. In this
situation the most interesting are the spontaneous transformations of
simple patterns, especially localized steady patterns --- autosolitons
(AS), into much more complicated ones. A lot of different types of these
transformations were recently observed in the experiments with
semiconductor and gas discharge structures
\cite{willebrand:cpp92,willebrand:pra92,niedernostheide:prb92}, chemical
reaction-diffusion systems \cite{lee:sci93,lee:pre95,lee:ln94}, and in
the numerical simulations
\cite{hagberg:prl94,hagberg:chaos94,elphick95}. At the same time, the
general theory of such transformations is absent.

The general asymptotic method of constructing the solutions in the form
of large-amplitude localized, periodic, and more complex one-dimensional
patterns, and the qualitative analysis of their stability in the systems
described by Eqs. (\ref{1}) and (\ref{2}) were developed by Kerner and
Osipov \cite{ko:book,ko:ufn89,ko:jetp78,ko:ftp79,ko:mk85,ko:jetp80,%
  ko:mk81,ko:jetp85,ko:jetp82}.  More formal and mathematically rigorous
analysis of one-dimensional patterns was carried out by Nishiura,
Mimura, and the co-authors \cite{mimura80,nishiura87,nishiura89}.
Static, pulsating, and traveling large-amplitude patterns in simple
reaction-diffusion systems were studied in detail
\cite{koga80,kko:mk84:1,kko:mk84:2,kko:dan84,ito92,ermentrout84,%
  klaasen84,kuo:pre95,rinzel73,casten75}.

In two and three dimensions Kerner and Osipov have constructed the
asymptotic solutions for radially-symmetric patterns, and also analyzed
the stability of one-dimensional patterns in higher dimensions
\cite{ko:book,ko:ufn89,ko:jetp80,ko:mk81,ko:jetp85,ko:jetp82}. Ohta,
Mimura, and Kobayashi have developed an approach which allowed them to
study the stability of one-dimensional and radially-symmetric patterns
in 2- and 3-dimensional versions of simple piecewise-linear model of
reaction-diffusion system \cite{ohta89}. This approach was further
developed by Petrich and Goldstein, who applied it to a version of the
FitzHugh-Nagumo model \cite{petrich94}.

In this paper we develop a systematic procedure of finding the
bifurcation points of an arbitrary $d$-dimensional pattern. Using this
procedure, we analyze the stability of the major types of patterns in
arbitrary dimensions. On the basis of this analysis, we draw conclusions
about possible scenarios of the transformations of patterns.

Our paper is organized as follows. In Sec. II we generalize the method
of constructing the asymptotic solution for one-dimensional and
radially-symmetric patterns developed in Refs.
\cite{ko:book,ko:ufn89,ko:jetp78,ko:ftp79,%
  ko:mk85,ko:jetp80,ko:mk81,ko:jetp85,ko:jetp82} to the case of
arbitrary $d$-dimensional patterns. In Sec. III we present the
derivation of the general dispersion relation governing the stability of
an arbitrary pattern. In Sec. IV we apply the obtained results to the
one-dimensional AS in higher dimensions. In Sec. V we analyze the
instabilities of the spherically-symmetric AS in three dimensions, and
cylindrically-symmetric AS in two and three dimensions. In Sec. VI we
summarize the obtained results and discuss their implications on the
evolution of patterns and also give comparisons with the experimental
and numerical data.

\section{asymptotic solutions for arbitrary d-dimensional patterns}

If we choose $L$ and $\tau_\eta$ as the units of length and time, we can
write Eqs. (\ref{1}) and (\ref{2}) as
\begin{equation}
\label{act}
\alpha {\frac{\partial \theta }{\partial t}}=\epsilon^2 \Delta \theta -
q(\theta,\eta,A),
\end{equation}
\begin{equation}
\label{inh}{\frac{\partial \eta }{\partial t}}=\Delta \eta -
Q(\theta,\eta,A),
\end{equation}
where $\epsilon \equiv l/L$ and $\alpha \equiv \tau_\theta / \tau_\eta$
are the ratios of the characteristic lengths and times of the activator
and the inhibitor, respectively. The boundary conditions for Eqs.
(\ref{act}) and (\ref{inh}) may be neutral or periodic.

Kerner and Osipov developed the qualitative theory of large-amplitude
patterns in reaction-diffusion systems
\cite{ko:jetp78,ko:ftp79,ko:mk85,%
  ko:jetp80,ko:mk81} (for a comprehensive review on the subject see
Refs. \cite{ko:book,ko:ufn89,ko:ufn90}).  They showed that the overall
type of patterns is determined by the values of $\epsilon $ and $\alpha
$, and the form of the nullcline of Eq. (\ref{act}), that is the
dependence $\eta (\theta )$ implicitly determined by equation $q(\theta
,\eta ,A)=0$ for a fixed value of $A$. For many of systems where
patterns may form this nullcline is N- or inverted N-like (Fig.
\ref{f1}).

According to the general qualitative theory \cite{ko:book,ko:ufn89},
when $\epsilon \ll 1$ and $\alpha \gg 1$, only static patterns may
form in the system; when $\alpha \ll 1$ and $\epsilon \gg 1$ only
traveling patterns may form; and when both $\epsilon \ll 1 $ and
$\alpha \ll 1$, all types of patterns --- static, traveling, and
pulsating --- may form.

From the mathematical point of view the fact that $\theta$ is activator
means that in some range of the system's parameters $q^{\prime}_\theta <
0$. In N-systems this condition is satisfied for $\theta_0 \lesssim
\theta \lesssim \theta_0^{\prime}$ (see Fig. \ref{f1}). The fact that
$\eta$ is inhibitor means that the following conditions hold
\cite{ko:book,ko:ufn89}
\begin{equation}
\label{inh:cond}Q^{\prime}_\eta > 0, ~~~~q^{\prime}_\eta
Q^{\prime}_\theta < 0,
\end{equation}
and in the whole range of the system's parameters the derivatives $%
Q^{\prime}_\theta$, $Q^{\prime}_\eta$, and $q^{\prime}_\eta$ do not
change signs.

The systems we are considering have a unique homogeneous state $\theta =
\theta_h$ and $\eta = \eta_h$, where $\theta_h$ and $\eta_h$ satisfy
\begin{equation}
\label{hom}q(\theta_h, \eta_h, A) = 0, ~~~~Q(\theta_h, \eta_h, A) = 0.
\end{equation}
The homogeneous state is stable for $A < A_0$, where $A_0$ is the point
where $\theta_h = \theta_0$ (see Fig. \ref{f1}) \cite{ko:book,ko:ufn89}.

As follows from the qualitative theory \cite{ko:book,ko:ufn89},
condition $\epsilon \ll 1$ is necessary for the existence of AS and
other large-amplitude patterns in the considered reaction-diffusion
systems. This fact allows to use $\epsilon$ as a natural small parameter
and construct asymptotic solutions by means of the singular perturbation
theory \cite{ko:book,ko:ufn89,ko:mk85,ko:jetp85}. Kerner and Osipov have
shown that as $\epsilon \rightarrow 0$, a pattern looks like a
collection of ``hot'' (high values of the activator) and ``cold'' (low
values of the activator) regions, separated by the walls whose width is
of the order of $\epsilon$
\cite{ko:book,ko:ufn89,ko:ftp79,ko:mk85,ko:jetp80,ko:mk81}. Thus, in the
limit $\epsilon \rightarrow 0$ any pattern in a $d$-dimensional N-system
can be described as a $(d-1)$-dimensional manifold ${\cal S}$,
corresponding to the walls of the pattern, which separates hot and cold
regions $\Omega_+$ and $\Omega_-$, respectively (Fig. \ref{domain}). In
general, ${\cal S}$ is a collection of an (infinite) number of
disconnected orientable submanifolds ${\cal S}_i$.

Let us introduce the orthogonal curvilinear coordinates around each
submanifolds ${\cal S}_i$. For a point $x$ let $\rho_i$ be the distance
from $x$ to ${\cal S}_i$, and the $d-1$-dimensional coordinate
$\vec{\xi}_i$ the projection to the submanifold $\rho_i = const$. The
value of $\rho_i$ is assumed to be positive, if $x \in \Omega_+$ and
negative otherwise. In the region of size $\sim \epsilon$ around each
${\cal S}_i$ the variation of the inhibitor in the direction
perpendicular to ${\cal S}_i$ and the variation of the activator along
${\cal S}_i$ are negligible compared to the variation of the activator
in the direction perpendicular to ${\cal S}_i$
\cite{ko:book,ko:ufn89,ohta89}. Therefore, in the vicinity of ${\cal
  S}_i$ stationary Eq. (\ref{act}) can be approximately written as
\begin{equation}
\label{sh}\epsilon^2 {\frac{d^2 \theta_i }{d \rho_i^2}} + \epsilon^2
K_i(\rho_i, \xi_i) {\frac{d \theta_i }{d \rho_i }} = q(\theta_i ,
\eta_s^i, A),
\end{equation}
where $K_i(\rho_i, \xi_i)$ is the curvature of ${\cal S}_i$ at a point
with the curvilinear coordinates $\rho_i$ and $\xi_i$. The boundary
conditions for $\theta_i$ in Eq. (\ref{sh}) are
\begin{equation}
\label{sh:bound}\theta_i(-\infty) = \theta_{s1}^i, ~~\theta_i(+\infty) = 
\theta_{s3}^i, ~~ \theta_i(0) = \theta_{s2}^i,
\end{equation}
where $\theta_{sk}^i$ satisfy $q(\theta_{sk}^i, \eta_s^i, A) = 0$, with
$\theta_{s1}^i < \theta_{s2}^i < \theta_{s3}^i$, and $\eta_s^i$
satisfies the consistency condition
\begin{equation}
\label{sh:cons}\epsilon^2 \int_{-\infty}^{+\infty} K_i(\rho_i, \xi_i)
\left( {\frac{ d \theta_i }{ d \rho_i }} \right)^2 d \rho_i =
  \int_{\theta_{s1}^i}^{\theta_{s3}^i} q(\theta, \eta_s^i, A) d \theta,
\end{equation}
which follows from Eq. (\ref{sh}), if we multiply it by $d \theta_i / d
\rho_i$ and integrate over $\rho_i$. Of course, the infinities in Eq.
(\ref {sh:bound}) actually mean that the boundary conditions for Eqs.
(\ref{sh}) should be satisfied sufficiently far from ${\cal S}$, namely
for $| \rho | \gg \epsilon$. Note that in the equations above $\xi_i$
appears only as a parameter.

When $K_i \rightarrow 0$ the solution of Eqs. (\ref{sh}) --
(\ref{sh:cons}) naturally transforms to the one-dimensional sharp
distribution (inner solution) \cite{ko:book,ko:ufn89,ko:ftp79,ko:mk85}.
\begin{equation}
\label{sh:1}\epsilon^2 {\frac{d^2 \theta_{sh} }{d \rho_i^2}} =
q(\theta_{sh}(\rho_i), \eta_s, A),
\end{equation}
where $\eta_s$ is a constant determined by equation
\begin{equation}
\label{eta:s}\int_{\theta_{s1}}^{\theta_{s3}} q(\theta, \eta_s, A)
d\theta = 0,
\end{equation}
and $\theta_{sk}$ are constants satisfying $q(\theta_{sk}, \eta_s, A) =
0$. The boundary conditions for Eq. (\ref{sh:1}) are given by Eq.
(\ref{sh:bound}) with $\eta_s^i = \eta_s$, $\theta_{sk}^i =
\theta_{sk}$. Thus, Eqs. (\ref{sh}) -- (\ref{sh:cons}) describe the
sharp distributions of the activator around ${\cal S}_i$ and take into
account the curvature of ${\cal S}_i$.

Far from ${\cal S}$ the characteristic length of the activator variation
is of order one, so for $|\rho_i| \gg \epsilon$ the solution of Eqs.
(\ref{act}) and (\ref{inh}) is approximately given by the smooth
distributions (outer solutions) $\theta_{sm}^\pm(x)$ and
$\eta_{sm}^\pm(x)$ defined for $x \in \Omega_\pm$, respectively, which
satisfy \cite{ko:book,ko:ufn89,ko:ftp79,ko:mk85,ko:jetp80,ko:mk81}
\begin{equation}
\label{sm}\Delta \eta_{sm}^\pm = Q( \theta_{sm}^\pm, \eta_{sm}^\pm, A),
~~~q(\theta_{sm}^\pm, \eta_{sm}^\pm, A) = 0
\end{equation}
for $x \in \Omega_\pm$, respectively, with the boundary conditions
\begin{equation}
\label{bc:sm}\eta_{sm}^\pm(x_i) = \eta_s^i,~~{\partial \eta_{sm}^+
  \over \partial \rho_i} = {\partial \eta_{sm}^- \over \partial
  \rho_i},~~\theta_{sm}^+(x_i) = \theta_{s3}^i,~~\theta_{sm}^-(x_i) =
\theta_{s1}^i
\end{equation}
for any $x_i \in {\cal S}_i$. Note that the shape of ${\cal S}$ itself
is determined self-consistently via Eqs. (\ref{sm}) and (\ref{bc:sm}).

According to the singular perturbations theory \cite{ko:book,ko:ufn89},
for $x \in \Omega_\pm $ the asymptotic solution of Eqs. (\ref{act}) and
(\ref{inh}) is given by the following composition of the sharp and
smooth distributions
\begin{equation}
\label{shsm}\theta^\pm(x)=\theta_{sm}^\pm(x)+ \sum_i (\theta_i(x) -
\theta_{s1,3}^i), ~\eta^\pm(x) = \eta_{sm}^\pm(x)
\end{equation}
where the sign ``plus'' goes with $\theta_{s3}^i$ and the sign ``minus''
goes with $\theta_{s1}^i$.

\section{general method for calculating instabilities}

Let us consider the problem linearized about the static solutions of
Eqs. (\ref{act}) and (\ref{inh}) with respect to the fluctuations
\begin{equation}
\label{fluc}\delta\theta(x,t)=\delta\theta(x) e^{i \omega
  t},~~~~~~~\delta\eta(x,t)=\delta\eta(x) e^{i \omega t}.
\end{equation}
The equations describing the fluctuations with the frequency $\omega$
will become
\begin{equation}
\label{tht:xy}i \alpha \omega \delta\theta = \epsilon^2 \Delta
\delta\theta - q^{\prime}_\theta ~(\theta(x),\eta(x))\delta\theta -
q^{\prime}_\eta(\theta(x), \eta(x))\delta\eta,
\end{equation}
\begin{equation}
\label{ett:xy}i \omega \delta\eta= \Delta \delta\eta -
Q^{\prime}_\theta(\theta(x),\eta(x))\delta\theta -
Q^{\prime}_\eta(\theta(x), \eta(x))\delta\eta.
\end{equation}
As was shown in the previous Section, the solutions $\theta(x)$ and
$\eta(x)$ in the form of a static pattern, around which Eqs. (\ref{act})
and (\ref{inh}) are linearized, are approximately given by Eqs.
(\ref{shsm}) for sufficiently small $\epsilon$.

According to the general qualitative theory \cite{ko:book,ko:ufn89}, the
stabilization of a pattern occurs due to the damping effect of the
inhibitor on the fluctuations of the activator. It was shown that only
those fluctuations of the activator that are localized in the walls of
the pattern and lead to their small displacements are dangerous for the
pattern's stability. To incorporate this fact into our analysis, let us
first consider the fluctuations in the vicinity of ${\cal S}_i$ with the
fixed distribution of the inhibitor. Putting $\delta\eta = 0$ in Eq.
(\ref{tht:xy}), we may write
\begin{equation}
\label{f:w}i \alpha \omega \delta\theta = - (\hat{H}^0_\theta +
\hat{H}_\theta^1 + \epsilon^2 \hat{S}_i ) \delta\theta ,
\end{equation}
where
\begin{equation}
\label{h:w}\hat{H}_\theta^0 = - \epsilon^2 {\frac{\partial ^2 }{\partial
    \rho_i^2}} + q^{\prime}_\theta( \theta_{sh}(\rho_i), \eta_s),
\end{equation}
\begin{equation}
\label{h1}\hat{H}_\theta^1 = - \epsilon^2 \hat{K} +
q^{\prime}_\theta(\theta(x), \eta(x)) -
q^{\prime}_\theta(\theta_{sh}(\rho_i), \eta_s),
\end{equation}
and the operator $\hat{S}_i$ is the part of the Laplacian acting on the
$d-1$-dimensional coordinates $\xi_i$, evaluated at ${\cal S}_i$ and
taken with the minus sign; $\hat{K}$ is the rest of the Laplacian
associated with the curvature of ${\cal S}_i$.

The lowest bound eigenstate of the operator $\hat{H}_\theta^0$ is
$\delta\theta_0 = d \theta_{sh} / d \rho_i$, which corresponds to the
eigenvalue $\lambda = 0$
\cite{ko:book,ko:ufn89,ko:ftp79,ko:mk85,ko:jetp80,ko:mk81}. Indeed, if
we differentiate Eq. (\ref{sh:1}) with respect to $\rho_i$, we will get
\begin{equation}
\label{h:0}\hat{H}_\theta^0 {\frac{d \theta_{sh} }{d\rho_i}} = 0.
\end{equation}
The function $\delta \theta_0 = d \theta_{sh} / d \rho_i$ has no nodes;
therefore, it corresponds to the lowest eigenstate. The eigenvalues
corresponding to the excited states are all of order one. This means
that the excited states, whose eigenvalues are all positive and of order
one, are highly damped and are therefore not important for our analysis
of the instabilities
\cite{ko:book,ko:ufn89,ko:ftp79,ko:mk85,ko:jetp80,ko:mk81}.

The operators $\hat{H}_\theta^1$ and $\epsilon^2 \hat{S}_i$ in Eq.
(\ref{f:w}) can be treated as perturbations to the operator
$\hat{H}_\theta^0$. Since the unperturbed operator $\hat{H}_\theta^0$
has no dependence on $\xi_i$, we need to introduce the orthogonal basis
of the states corresponding to the surface modes on ${\cal S}_i$. As
such a basis, we may choose the eigenfunctions of the operator
$\hat{S}_i$.  Then the dangerous fluctuations $\delta\theta_{sh}$ are
linear combinations of the functions $\delta\theta_{il}$, where
\begin{equation}
\label{h:w:ef}\delta\theta_{il} = \phi_l(\xi_i) {\frac{d \theta_{sh} }{d 
    \rho_i}},~~~~\hat{S}_i \phi_l = \nu_{li} \phi_l,
\end{equation}
and $\phi_l$ satisfy
\begin{equation}
\label{phi:ort}\int_{{\cal S}_i} \phi_l^* (\xi_i) \phi_l (\xi_i) d^{d-1}
\xi_i = \delta_{l l^{\prime}}.
\end{equation}
Of course, for each ${\cal S}_i$ there is its own set of $\phi_l$. We
will frequently omit the indices such as $i$ and $\omega$, wherever it
does not lead to ambiguities, in order to simplify the notation.

Up to now we ignored the reaction of the inhibitor on the fluctuations
of the activator. According to the general qualitative theory, this
reaction can be included into our analysis by means of the perturbation
theory \cite{ko:book,ko:ufn89}. The main problem here is to correctly
find the response of the inhibitor on the dangerous fluctuations of the
activator. The formal solution of the problem is of no practical use
since one has to diagonalize a complicated operator, nor is the
expansion in the eigenfunctions since one has to consider the whole
spectrum of the problem. A way to do this is to use the idea of singular
perturbation theory and separate $\delta\theta$ into the localized
(sharp) and the delocalized (smooth) parts (for more rigorous derivation
see Appendix \ref{eigen}):
\begin{equation}
\label{LE}\delta\theta=\delta\theta_{sh}+\delta\theta_{sm}.
\end{equation}
Far from ${\cal S}$ Eq. (\ref{tht:xy}) will become
\begin{equation}
\label{lc:xy}i \alpha \omega \delta\theta_{sm} = -(q^{\prime}_\theta)_{sm} 
\delta\theta_{sm} - (q^{\prime}_\eta)_{sm} \delta\eta,
\end{equation}
where the subscript ``sm'' means that the derivatives are evaluated at
the smooth distributions, that is
\begin{equation}
\label{q:sm}(q^{\prime}_\theta)_{sm}=
q^{\prime}_\theta~(\theta_{sm}(x),\eta_{sm}(x)),~~(q^{\prime}_\eta)_{sm}
= q^{\prime}_\eta (\theta_{sm}(x), \eta_{sm}(x)).
\end{equation}
As follows from the general qualitative theory \cite{ko:book,ko:ufn89},
for all types of instabilities the condition $\alpha \omega \ll 1$
is satisfied. For this reason we may neglect the left-hand side in Eq.
(\ref{lc:xy}), and obtain that
\begin{equation}
\label{lc:ko}\delta\theta_{sm}= - {\frac{(q^{\prime}_\eta)_{sm} }
  {(q^{\prime}_\theta)_{sm}}} ~\delta\eta
\end{equation}

Let us substitute Eq. (\ref{LE}) into Eqs. (\ref{tht:xy}) and
(\ref{ett:xy}). Using Eqs. (\ref{h:0}) and (\ref{lc:ko}), we can rewrite
Eq. (\ref{tht:xy}) around ${\cal S}$, and Eq. (\ref{ett:xy}) in the
whole space as
\begin{equation}
\label{tht}(i \alpha \omega + \epsilon^2 \hat{S}_i + \hat{H}_\theta^1 )
\delta\theta_{sh} = - q^{\prime}_\eta \delta\eta,
\end{equation}
\begin{equation}
\label{ett}\left[i\omega - \Delta + (Q^{\prime}_\eta)_{sm} - {\frac{%
    (q^{\prime}_\eta)_{sm} (Q^{\prime}_\theta)_{sm}
    }{(q^{\prime}_\theta)_{sm} }} \right] \delta\eta = -
Q^{\prime}_\theta \delta\theta_{sh}.
\end{equation}
Note that we neglected the term $\hat{H}_\theta^0 \delta \theta_{sm}$ in
the left-hand side of Eq. (\ref{tht}) since it does not contribute in
the first order of the perturbation theory. Also note that in writing
Eq.  (\ref{ett}) we replaced the true distributions $\theta(x)$ by the
smooth distributions $\theta_{sm}(x)$. It is easy to see that this
replacement gives negligible difference in $\delta\eta$.

Let us solve Eq. (\ref{ett}) for $\delta\eta$ by means of the Green's
function
\begin{equation}
\label{Green}\delta\eta(x) = - \int Q^{\prime}_\theta(x^{\prime}) G(x,
x^{\prime}) \delta\theta_{sh} (x^{\prime}) dx^{\prime},
\end{equation}
where $G(x, x^{\prime})$ satisfies
\begin{equation}
\label{G}\left[ i \omega - \Delta + C + V(x) \right] G(x, x^{\prime}) = 
\delta(x - x^{\prime}),
\end{equation}
where
\begin{equation}
\label{C}C = Q^{\prime}_\eta(\theta_h, \eta_h) - {\frac{q^{\prime}_\eta 
    (\theta_h, \eta_h) Q^{\prime}_\theta(\theta_h, \eta_h)
    }{q^{\prime}_\theta (\theta_h, \eta_h) }},
\end{equation}
and
\begin{equation}
\label{V}V(x) = \left[ (Q^{\prime}_\eta)_{sm} - {\frac{
    (q^{\prime}_\eta)_{sm} (Q^{\prime}_\theta)_{sm}
    }{(q^{\prime}_\theta)_{sm} }} \right] - C.
\end{equation}

In view of Eq. (\ref{Green}), the right-hand side of Eq. (\ref{tht}) can
be regarded as an operator $\hat{R}$ acting on $\delta \theta_{sh}$
\begin{equation}
\label{R}\hat{R}[\delta\theta_{sh}]= q^{\prime}_\eta(x) \int
Q^{\prime}_\theta(x^{\prime}) G(x,x^{\prime})
\delta\theta_{sh}(x^{\prime}) d^d x^{\prime}.
\end{equation}
Since the sharp fluctuation $\delta\theta_{sh}$ is the linear
combination of the functions $\delta\theta_{il}$ defined in Eq.
(\ref{h:w:ef}), the integral in Eq. (\ref{R}) can be easily calculated.
Taking into account that $d \theta_{sh} / d \rho_i$ are close to
delta-functions \cite{ko:book,ko:ufn89}, we may write
\begin{equation}
\label{R:calc}\hat{R}[ \delta\theta_{il}] = \left( Q(\theta_{s3},
\eta_s) - Q(\theta_{s1}, \eta_s) \right) q^{\prime}_\eta(x) \int_{{\cal
    S}_i} G(x, \xi_i) \phi_l^i( \xi_i) d^{d-1} \xi_i,
\end{equation}
where $\xi_i$ denotes the point on ${\cal S}_i$ and the integration is
over ${\cal S}_i$. The matrix elements of $\hat{R}$ can be calculated
analogously. The result of the calculation is
\begin{equation}
\label{R:me}\langle i^{\prime}l^{\prime}| \hat{R} | i l \rangle = - B
\int_{{\cal S}_{i^{\prime}}} \int_{{\cal S}_{i}} G( \xi_{i^{\prime}},
\xi_i) \phi^*_{l^{\prime}}(\xi_{i^{\prime}}) \phi_l (\xi_{i}) d^{d-1}
\xi_{i^{\prime}} d^{d-1} \xi_{i},
\end{equation}
where
\begin{equation}
\label{B}B = - (Q(\theta_{s3}, \eta_s) - Q(\theta_{s1}, \eta_s))
\int_{\theta_{s1}}^{\theta_{s3}} q^{\prime}_\eta( \theta, \eta_s) d
\theta.
\end{equation}
is a constant depending on $A$ only. Note that in accordance with Eq.
(\ref {inh:cond}) the value of $B$ is positive.

When the distance between the different ${\cal S}_i$ is much greater
than $\epsilon$, the overlap between the different $\delta\theta_{il}$
is negligible, so the operator $\hat{H}_\theta^1$ is diagonal in the
$i$-indices. Then, in the first order of the perturbation theory Eqs.
(\ref{tht}), (\ref{R}), and (\ref{R:me}) reduce to
\begin{equation}
\label{disp:gen}(i \alpha \omega + \epsilon^2 \nu_{li})
\delta_{ii^{\prime}} \delta_{l l^{\prime}} = \epsilon Z^{-1} [ \langle
i^{\prime}l^{\prime}| \hat{R} | i l \rangle - \delta_{ii^{\prime}}
\langle i l ^{\prime}| \hat{H}_\theta^1 | i l \rangle ],
\end{equation}
where
\begin{equation}
\label{Z}Z = \epsilon \int_{-\infty}^{+\infty} \left( {\frac{d
    \theta_{sh} }{d \rho}} \right)^2 d \rho.
\end{equation}
Note that the value of $Z$ is of order one since the characteristic
length of the activator variation is $\epsilon$.

Eq. (\ref{disp:gen}) is the principal equation which determines the
stability of an arbitrary $d$-dimensional pattern in N-systems. This
equation was derived with an accuracy to $\epsilon \ll 1$ and
$\epsilon K_{max} \ll 1$, where $K_{max}$ is the maximum curvature
of a given pattern.

If a pattern possesses certain symmetries, the operators $\hat{R}$ and
$\hat{H}_\theta^1$ are diagonal in the $l$-indices.  In this case the
operators in Eq.  (\ref{disp:gen}) can be easily diagonalized (see the
following sections).  The main problem is to find the Green's function
$G(x, x^{\prime})$. Once this is done, we can obtain the ``dispersion
relation'', which relates $\omega$ with the values of $A$, $\epsilon$
and $\alpha$ for different types of fluctuations.

\section{static one-dimensional autosoliton in two and three dimensions}

Let us apply the procedure developed in the previous section to the
simplest pattern --- static one-dimensional autosoliton in two or three
dimensions (Fig. \ref{as}). Since the AS is localized, the distributions
of the activator and the inhibitor on its periphery go to the stable
homogeneous state $\theta = \theta_h$ and $\eta = \eta_h$, where
$\theta_h$ and $\eta_h$ are determined by Eq. (\ref{hom}).  In this
case, according to Eqs. (\ref{inh:cond}) and (\ref{C}), the value of $C
> 0$ since for $\theta_h < \theta_0$ the value of $q'_\theta( \theta_h,
\eta_h) > 0$ \cite{ko:book,ko:ufn89}.

For a one-dimensional AS the manifold ${\cal S}$ consists of two
parallel planes where the AS walls are localized. We can choose the
coordinate directions in such a way that these planes are perpendicular
to the $z$-axis. Then the solution for the AS will depend only on $z$.

Since the considered static 1-d AS is symmetric with respect to its
center \cite{ko:book,ko:ufn89}, we can assume that the positions of its
left and its right walls are $z_1 = - {\cal L}_s /2$ and $z_2 = {\cal
  L}_s /2$, respectively, where ${\cal L}_s$ is the distance between the
walls. Note that the value of ${\cal L}_s$ itself can be used as a
bifurcation parameter instead of $A$, since there is a one-to-one
correspondence between them in the whole region of AS existence
\cite{ko:book,ko:ufn89}. For this reason, here and further we will use
the AS width ${\cal L}_s$ as the bifurcation parameter instead of $A$.

Since the ${\cal S}_{1,2}$ are flat, the curvilinear coordinates
$\rho_{1,2}$ coincide with $z - z_1$ and $-(z - z_2)$, respectively (the
signs are consistent with our definition of $\rho_i$ given in Sec. II),
and the coordinates $\xi_i$ coincide with the rest of the coordinates of
space.  Of course, $\hat{K} \equiv 0$.

According to our procedure, let us first look at the operator
$\hat{S}_i$.  Here $\hat{S}_{1,2} = - \Delta_{\xi_{1,2}}$ is the
Laplacian, acting on the local coordinates on ${\cal S}_{1,2}$. The
eigenfunctions of this operator are just plane waves with the wavevector
$k$ along ${\cal S}_{1,2}$, whose eigenvalues are $\nu_k = k^2$.

If we substitute the eigenfunctions of $\hat{S}_{1,2}$ into Eq.
(\ref{R:me}) and use the fact that the system has translational
invariance in the $\xi$-directions, by integrating over $\xi_i$-s we
will get
\begin{eqnarray} \label{R:ko}
  \langle 1, k | \hat{R} | 1, k' \rangle = - B G_{k}\left(\frac{{\cal
      L}_s}{2}, \frac{{\cal L}_s}{2} \right) \delta( k - k' ), \\ 
  \langle 1, k | \hat{R} | 2, k'\rangle = - B G_{k} \left(-\frac{{\cal
      L}_s}{2}, \frac{{\cal L}_s}{2} \right) \delta( k - k'),
\end{eqnarray}
where $G_k(z, z')$ is the Fourier-transform of $G(x,x')$ in $\xi$.
Because the AS is symmetric with respect to its center, the values of
the matrix elements satisfy $\langle 1, k | \hat{R} | 1, k \rangle =
\langle 2, k | \hat{R} | 2, k \rangle$ and $\langle 1, k | \hat{R} | 2,
k \rangle = \langle 2, k | \hat{R} | 1, k \rangle $. Thus, the matrix
elements of the operator $\hat{R}$ in this case can be expressed in
terms of the values of the Fourier-transformed Green's function at the
particular points of the $z$-axis.

Let us now turn to the operator $\hat{H}_\theta^1$. As was shown in the
general qualitative theory \cite{ko:book,ko:ufn89}, the functions $d
\theta_{sh}/ d \rho_{1,2}$ decay exponentially at distances much larger
than $\epsilon$. Because of this, for ${\cal L}_s \gg \epsilon \log
(\epsilon^{-1})$ the overlap of these functions can be neglected.
Following the notation of Refs.  \cite{ko:book,ko:ufn89}, we write the
diagonal elements of $\hat{H}_\theta^1$ as
\begin{equation}
\label{lambda0}\langle 1, k| \hat{H}_\theta^1 | 1, k^{\prime}\rangle =
\langle 2, k| \hat{H}_\theta^1 | 2, k^{\prime}\rangle = \epsilon^{-1} Z
\lambda_0 \delta( k - k^{\prime}),
\end{equation}
where $\lambda_0$ is of order $\epsilon$.  The operator in the
right-hand side of Eq. (\ref{disp:gen}) can be trivially diagonalized.
Since the AS possesses central symmetry, the eigenstates of this
operator are the symmetric and the antisymmetric combinations of
$\delta\theta_{1k}$ and $\delta\theta_{2k}$. These states correspond to
the symmetric and the antisymmetric deformations of the AS walls.  As a
result, introducing the functions
\begin{eqnarray}
  R_0(k, \omega) = G_k\left(\frac{{\cal L}_s}{2}, \frac{{\cal L}_s}{2}
\right) + G_k \left(\frac{{\cal L}_s}{2}, - \frac{{\cal L}_s}{2} \right)
, \label{R0}
\\
R_1(k, \omega) = G_k\left(\frac{{\cal L}_s}{2}, \frac{{\cal
    L}_s}{2}\right) - G_k\left(\frac{{\cal L}_s}{2}, -\frac{{\cal
    L}_s}{2} \right) \label{R1},
\end{eqnarray}
from Eq. (\ref{disp:gen}) we obtain the dispersion relation
\begin{equation}
\label{ko}i \alpha \omega + \epsilon^2 k^2 + \lambda_0 = - \epsilon B
Z^{-1} R_{0,1}(k, \omega).
\end{equation}
Here the subscript ``0'' corresponds to the symmetric, and the subscript
``1'' corresponds to the antisymmetric modes of fluctuations. The value
of $\lambda_0$ can be calculated indirectly. Since the AS possesses
translational invariance in the $z$-direction, Eq. (\ref{ko}) should be
identically satisfied for $k = 0$ and $\omega = 0$ for the antisymmetric
fluctuation.  This immediately means that
\begin{equation}
\label{lambda0:calc}\lambda_0 = - \epsilon B Z^{-1} R_1(0, 0).
\end{equation}

Eqs. (\ref{R0}) and (\ref{R1}) define the functions $R_0(k, \omega)$ and
$R_1(k, \omega)$, which describe the inhibitor reaction on the dangerous
fluctuations of the activator. In general the potential $V(z)$ in Eq.
(\ref{G}) which determines the Green's function is some non-trivial
function of $z$, so $R_{0,1}( k, \omega)$ are some complicated functions
of ${\cal L}_s$. In the case of piecewise-linear model one can calculate
the values of $R_0(k, \omega)$ and $R_1(k, \omega)$ explicitly (see
Appendix \ref{p-l}). One can see that the dispersion relations for the
fluctuations obtained in this case are identical to those found earlier
by Ohta, Mimura, and Kobayashi in Ref. \cite{ohta89} who used a
different approach.

To calculate the critical values of $A$ and the parameters of the
critical fluctuations in general, we need to know the detailed form of
the Green's function. According to Eq. (\ref{G}), the Fourier
transformed Green's function $G_k(z, z')$ is governed by
\begin{equation}
\label{G:ko}\left[ - {\frac{\partial ^2 }{\partial z^2}} + i \omega +
k^2 + C + V(z) \right] G_k(z, z^{\prime}) = \delta( z - z^{\prime}).
\end{equation}
As was said earlier, the potential in the operator in the left-hand side
of Eq. (\ref{G:ko}) is some complicated function of $z$. However, the
problem is greatly simplified for finding instabilities, since, as was
shown in the general qualitative theory \cite{ko:book,ko:ufn89}, most of
the instabilities occur when ${\cal L}_s \ll 1$ (see also the
results below). This allows to use the value of ${\cal L}_s$ as a small
parameter and expand the functions $R_{0,1}(k, \omega)$ in terms of it.

Regarding all this, we are now able to construct the perturbation
expansion for the Green's function, considering the potential $V(z)$ in
Eq. (\ref{G:ko}) as a perturbation. The unperturbed Green's function
satisfies
\begin{equation}
\label{G:koc}\left[ -{\frac{\partial ^2 }{\partial z^2}} + k^2 + i
\omega + C \right] G_k^{(0)}(z, z^{\prime}) = \delta (z - z^{\prime}),
\end{equation}
The solution of Eq. (\ref{G:koc}) is well known
\begin{equation}
\label{G0}G^{(0)}_k(z , z^{\prime}) = \frac{\exp \{ -\sqrt{C + k^2 + i
    \omega}~ | z - z^{\prime}| \} }{2 \sqrt{C + k^2 + i \omega}}.
\end{equation}
To calculate the corrections to the Green's function, we use a formula
\begin{equation}
\label{Gn}G^{(n)}_k (z, z^{\prime}) = - \int_{-\infty}^{+\infty}
V(z^{\prime\prime}) G^{(0)}_k(z , z^{\prime\prime})
G^{(n-1)}_k(z^{\prime\prime}, z^{\prime}) dz^{\prime\prime},
\end{equation}
where $G^{(n)}_k(z, z^{\prime})$ is the $n$-th correction.

Let us denote the contributions from $G^{(n)}_k$ to $R_{0,1}(k, \omega)$
as $R^{(n)}_{0,1}(k, \omega)$, respectively. Since for small values of
${\cal L}_s$ the coefficients $B$, $C$, and $Z$ only weakly depend on
$A$, we may replace them by their values at $A = A_b$, where the AS size
becomes formally zero when $\epsilon \rightarrow 0$
\cite{ko:book,ko:ufn89}.  Then the functions $R_0^{(0)}(k, \omega)$ and
$R_1^{(0)}(k, \omega)$ can be written as
\begin{equation}
\label{R0:ko}R_0^{(0)}(k, \omega) = {\frac{ 1 }{2 \sqrt{C + k^2 + i
      \omega} }} \left\{ 1 + \exp (-{\cal L}_s \sqrt{C + k^2 + i \omega}
) \right\},
\end{equation}
\begin{equation}
\label{R1:ko}R_1^{(0)}(k, \omega) = {\frac{{\cal L}_s }{2}} -
{\frac{{\cal L}_s^2 \sqrt{C + k^2 + i \omega} }{4}} + {\cal O}({\cal
  L}_s^3).
\end{equation}
Note that $R_0^{(n)} = {\cal O} ({\cal L}_s^n)$, and because of the
central symmetry of the potential $R_1^{(n)} = {\cal O}({\cal
  L}_s^{2n+1})$.

Substituting these values of $R_{0,1}(k, \omega)$ into Eq. (\ref
{lambda0:calc}), we will obtain that the leading term of $\lambda_0$ is
\begin{equation}
\label{lambda0:LS}\lambda_0 = - \frac{\epsilon Z^{-1} B {\cal L}_s}{2}.
\end{equation}

Having calculated the value of $\lambda_0$, we may write the dispersion
relation for the symmetric fluctuations
\begin{equation}
\label{ko:symm}i \alpha \omega + \epsilon^2 k^2 = \epsilon B Z^{-1}
\left\{ {\frac{{\cal L}_s }{2}} - {\frac{1 }{2 \sqrt{C + k^2 + i \omega}
      }} \{1 + \exp(- {\cal L}_s \sqrt{C + k^2 + i \omega}) \} -
  R^{(1)}_0(k, \omega) \right\} + {\cal O}({\cal L}_s^2).
\end{equation}

As was shown in the general qualitative theory
\cite{ko:book,ko:ufn89,ko:jetp80,ko:mk81}, when $\alpha$ is big enough,
for ${\cal L}_s > {\cal L}_{c1}$ the one-dimensional AS becomes unstable
with respect to the fluctuation with $Re~\omega = 0$ and $k = k_c \gg
1$, corresponding to the corrugation of the AS walls. Note that,
according to Eq. (\ref{Gn}), for $k_c \gg 1$ the value of $R_0^{(1)}(k,
\omega)$ which is of order ${\cal L}_s$, contains a small factor
$\propto k^{-2}$, so it can be neglected. Let us calculate the values of
${\cal L}_{c1}$ and $k_c$.  Putting $\omega = 0$ and neglecting $C$ in
comparison with $k^2$ in Eq.  (\ref {ko:symm}), we will get a
transcendent equation, which can be solved for small values of ${\cal
  L}_s$. The result is
\begin{equation}
\label{corr}k_c = 0.71 \left( {\frac{\epsilon Z }{B }} \right)^{-1/3},
~~~ {\cal L}_{c1} = 2.64 \left( {\frac{\epsilon Z }{B}} \right)^{1/3}.
\end{equation}
Note that the dependences of ${\cal L}_{c1}$ and $k_c$ on $\epsilon$
coincide with those obtained in the qualitative theory for ${\cal L}_s
\ll 1$ \cite{ko:book,ko:ufn89}.

In the case $\alpha \ll \epsilon$ there is an instability for
${\cal L}_s > {\cal L}_\omega$ with respect to the fluctuation
describing the pulsations of the AS with the frequency $\omega =
\omega_c \gg 1$, and $k = 0$
\cite{ko:book,ko:ufn89,ko:jetp85,ko:jetp82}.  As before, the term
$R_0^{(1)}(k, \omega)$ contains the small factor $\omega^{-1}$ and can
be neglected.  Solving the transcendent equation obtained in this case,
we have
\begin{equation}
\label{puls}\omega_c = 0.73 \left( {\frac{\alpha Z }{\epsilon B}}
\right)^{-2/3}, ~~~~{\cal L}_\omega = 0.96 \left( {\frac{ \alpha Z }
  {\epsilon B}} \right)^{1/3}.
\end{equation}
Again, the calculated dependences of ${\cal L}_\omega$ and $\omega_c$ on
the ratio $\alpha / \epsilon$ coincide with those obtained in the
qualitative theory \cite{ko:book,ko:ufn89}, and in the analytic studies
of the piecewise-linear model \cite{kuo:pre95}.  Also, comparing ${\cal
  L}_{c1}$ and ${\cal L}_\omega$, one can see that the pulsating
instability occurs before the corrugation instability if $\alpha <
21\epsilon^2$.

When the size of the AS is greater than the critical size determined by
Eq. (\ref{corr}), the increment of the growing fluctuations may be very
small. Indeed, according to Eq. (\ref{ko:symm}), for ${\cal L}_s \gg
{\cal L}_{c1}$ we obtain that the increment of the most dangerous
fluctuations is $\gamma \simeq \epsilon B {\cal L}_s /(2 \alpha Z)$.

Now let us turn to the antisymmetric fluctuations. According to Eq.
(\ref{R1:ko}), the first term in $R_1(k, \omega)$ which depends on
$\omega$ and $k$ is of order ${\cal L}_s^2$. As was mentioned earlier,
the first correction $R_1^{(1)}(k, \omega)$ is of the order ${\cal
  L}_s^3$, so it can be neglected. Then Eq. (\ref{ko}) for the
antisymmetric fluctuations can be written as
\begin{equation}
\label{ko:anti}i \alpha \omega + \epsilon^2 k^2 = {\frac{ \epsilon B
    Z^{-1} {\cal L}_s^2 }{4}} \left\{ \sqrt{C + k^2 + i \omega} -
\sqrt{C} \right\} + {\cal O}({\cal L}_s^3).
\end{equation}

According to the general qualitative theory \cite{ko:book,ko:ufn89},
there are two types of antisymmetric instabilities: wriggling of the AS
walls, and formation of a traveling AS. The first instability is
realized when $Re~\omega = 0$, $k \rightarrow 0$ and ${\cal L}_s > {\cal
  L}_{c2}$. The second is realized when $\alpha \ll 1$, $k = 0$,
and ${\cal L}_s > {\cal L}_T$. We may expand the the right-hand side of
Eq.  (\ref{ko:anti}) in the powers of $k$ and $\omega$ and obtain the
following expression
\begin{equation}
\label{ko:anti0}i \alpha \omega + \epsilon^2 k^2 = \epsilon b {\cal
  L}_s^2 (k^2 + i \omega),
\end{equation}
where we retained only the first nonvanishing terms. The constant $b$ is
given by
\begin{equation}
\label{b}b = {\frac{B }{8 Z C^{1/2} }}.
\end{equation}
One can see from Eq. (\ref{ko:anti0}) that the instability $Im~\omega <
0$ occurs at $k = 0$ for ${\cal L}_s > {\cal L}_T$, where
\begin{equation}
\label{Lt}{\cal L}_T = \left( {\frac{\alpha }{b \epsilon}} \right)^{1/2},
\end{equation}
or for $Re~\omega = 0$ and ${\cal L}_s > {\cal L}_{c2}$, where
\begin{equation}
\label{Lc2}{\cal L}_{c2} = \left( {\frac{\epsilon }{b}} \right)^{1/2}.
\end{equation}
Comparing these two formulas, one can see that traveling AS forms before
the walls of the static AS become unstable with respect to wriggling,
when $\alpha < \epsilon^2$. This fact is also an obvious consequence of
an additional symmetry present in Eqs. (\ref{act}) and (\ref{inh}) at
$\alpha = \epsilon^2$.

Although for ${\cal L}_s > {\cal L}_{c2}$ the AS is unstable, the
increment of the most dangerous fluctuations may be extremely small.
Indeed, according to Eq. (\ref{ko:anti}), for ${\cal L}_{c2} \ll 
{\cal L}_T$ and ${\cal L}_{c2} \ll {\cal L}_s \ll {\cal L}_{c1}$ we have 
\begin{equation}
\label{incr}\gamma_{max} = \alpha^{-1} \left(
{\frac{ {\cal L}_s^2 B }{8 Z }} \right)^2 \ll 1 ~~{\rm for}~~k =
k_{max} = {\frac{{\cal L}_s^2 B } {8 \epsilon Z}} \gg 1.
\end{equation}

Comparison of the expressions in Eqs. (\ref{corr}) and (\ref{Lc2}) for
the critical values of ${\cal L}_s$ for symmetric and antisymmetric
fluctuations shows that for $\epsilon \ll 1$ the wriggling
instability always emerges before the corrugation instability. Likewise,
according to Eqs. (\ref{puls}) and (\ref{Lt}), for sufficiently small
ratios $\alpha / \epsilon$ traveling AS forms before the pulsating one.
Notice that these general conclusions are in agreement with the analytic
investigation of AS in the piecewise-linear model
\cite{ohta89,kuo:pre95}.

When the size of the AS becomes comparable with $\epsilon$, the overlap
between the eigenfunctions of the operator $\hat{H}_\theta^0$ becomes
significant, what leads to the instability of the AS. As was shown in
the general qualitative theory \cite{ko:book,ko:ufn89}, the instability
of an AS with ${\cal L}_s \sim \epsilon$ occurs with respect to
symmetric fluctuations. Since the asymptotic behavior of the sharp
solutions is exponential, an extra piece in Eq. (\ref{ko:symm}) from the
operator $\hat{H}^0_\theta$ will have the form $a \exp ( - {\cal L}_s /
\tilde{l})$, where $\tilde{l} = \epsilon q^{\prime}_\theta(\theta_{s3},
\eta_s)^{-1/2}$, $a$ is some constant of order 1
\cite{ko:book,ko:ufn89,ko:ftp79,ko:mk85,ko:jetp80,ko:mk81}.  Then we
obtain that for $\alpha \gg \epsilon$ the instability occurs with
respect to the fluctuations with $Re~\omega = 0$ and
\begin{equation}
\label{Lb}k = k_c = 2^{-1/3} \left( {\frac{\epsilon Z }{B }}
\right)^{-1/3}, ~~~{\cal L}_s < {\cal L}_{cb} = \frac{4 \tilde{l}}{3}
\log \epsilon^{-1},
\end{equation}
whereas for $\alpha \ll \epsilon$ the instability is realized with
respect to the fluctuations with $k = 0$ and
\begin{equation}
\label{Lbom}\omega = \omega_c = 2^{-1/3} \left( {\frac{\alpha Z
    }{\epsilon B }} \right)^{-2/3},~~~ {\cal L}_s < {\cal L}_{b\omega} =
- \frac{\tilde{l}}{3} \log \alpha \epsilon^2.
\end{equation}
Eqs. (\ref{Lb}) and (\ref{Lbom}) were calculated with the logarithmic
accuracy and coincide with those obtained in the qualitative theory
\cite{ko:book,ko:ufn89}.

In one-dimensional systems with $\alpha \gg \epsilon$ the value of
${\cal L}_b$ at which the AS disappears will be slightly different from
${\cal L}_{cb}$, since the only possible value for $k$ there is zero.
Putting $k = 0$ in the dispersion relation we obtain that ${\cal L}_b =
\tilde{l} \log \epsilon^{-1}$ \cite{ko:book,ko:ufn89}.

Up to now we considered the AS whose width ${\cal L}_s \ll 1$. This
is justified in 2- or 3-dimensional systems since, according to Eqs.
(\ref{corr}) and (\ref{Lc2}), the AS becomes unstable when ${\cal L}_s
\ll 1$. In one dimension, however, an AS remains stable up to the
values of ${\cal L}_s \sim 1$, if $\alpha \gg \epsilon$. In this
situation, according to the general qualitative theory
\cite{ko:book,ko:ufn89}, there is another effect causing the
instability. Namely, when ${\cal L}_s$ is close to some ${\cal L}_d \sim
1$ at which the activator in the AS center reaches the value of
$\theta_0^{\prime}$, such that $q^{\prime}_\theta( \theta_0^{\prime},
\eta_0^{\prime}) = 0$, the solution in the form of an AS disappears. As
a result, a local breakdown occurs at the AS center, causing the AS to
split, so that eventually the whole system becomes filled with a complex
pattern \cite{ko:book,ko:ufn89}.  This can also be seen from Eq.
(\ref{ko}), if we take into account that at the point $z = 0$ where
$\theta = \theta_0^{\prime}$ the potential $V(z)$ in Eq. (\ref{G:ko})
becomes singular. In the case $\alpha \lesssim \epsilon$ the values of
${\cal L}_\omega$ and ${\cal L}_T$ corresponding to the instabilities
leading to formation of pulsating and traveling AS, respectively, may be
less than ${\cal L}_d$, so the AS cannot reach the point where the local
breakdown occurs. There is also a possibility that the point $A = A_d$
where ${\cal L}_s = {\cal L}_d$ is preceded by the point $A = A_c$ where
the homogeneous state of the system becomes unstable
\cite{ko:book,ko:ufn89}.  In this case the periphery of the AS becomes
unstable. This can also be seen from Eq.  (\ref{ko}) if we take into
account that, according to Eq. (\ref{G:ko}), for $A > A_0$ the value of
$C$ becomes negative and therefore the tails of the Green's function
will become oscillatory. When $\alpha$ gets larger, the right-hand side
of Eq. (\ref{ko}) always remains of order one, so at some critical
values of $\alpha \gtrsim \epsilon$ the instability leading to the
formation of traveling and pulsating AS disappears.

\section{higher-dimensional radially-symmetric autosolitons}

Let us now turn to higher-dimensional AS. First we consider spherically
symmetric AS in three dimensions. In this case there is only one
manifold ${\cal S}$ which is simply a sphere of radius ${\cal R}_s$. As
a set of orthogonal curvilinear coordinates we choose the usual
spherical coordinates $\rho$, $\vartheta$ and $\varphi$, except $\rho$
will be measured from the surface of the sphere rather than from the
origin, in order to be consistent with our initial definition.

The operator $\hat{S}$ in the the considered case is
\begin{equation}
\label{S:3}
\hat{S} = - {\cal R}_s^{-2} \left\{ {\frac{ 1 }{\sin \vartheta}}
{\frac{\partial }{\partial \vartheta }} \sin \vartheta {\frac{\partial }
  {\partial \vartheta }} + {\frac{1 }{\sin^2 \vartheta }}
{\frac{\partial ^2 }{\partial \varphi^2}} \right\}.
\end{equation}
The eigenfunctions of the operator $\hat{S}$ are just spherical
harmonics $\phi_{lm} = Y_{lm}(\vartheta, \varphi) / {\cal R}_s$ with the
eigenvalues $\nu_l = l(l + 1) / {\cal R}_s^2$. The factor $1/ {\cal
  R}_s$ in the eigenfunction ensures the proper normalization.

Now let us calculate the matrix elements. First of all, since the system
possesses spherical symmetry, the only nonvanishing matrix elements of
the operator $\hat{H}_\theta^1$ are the diagonal elements, which are all
equal to each other. Due to the same symmetry, the operator $\hat{R}$ is
also diagonal, and its diagonal elements are independent of $m$.
According to Eq.  (\ref{R:me}), the diagonal matrix elements of
$\hat{R}$ are
\begin{equation}
\label{Rme:3}\langle l m | \hat{R} | l m \rangle \equiv - B
R_{l}(\omega),
\end{equation}
where
\begin{equation}
\label{R:3}R_{l}(\omega) = \int \int G({\cal R}_s, \vartheta, \varphi;
{\cal R}_s, \vartheta^{\prime}, \varphi^{\prime}) Y_{lm}^*(\vartheta,
\varphi) Y_{lm}(\vartheta^{\prime}, \varphi^{\prime}) {\cal R}_s^2
do^{\prime}do,
\end{equation}
$do$ is the element of the solid angle, and the Green's function is
written in terms of the spherical coordinates.

To calculate $R_l(\omega)$ let us note that $R_l(\omega) = {\cal R}_s^2
G_l( {\cal R}_s, {\cal R}_s)$, where
\begin{equation}
\label{Gl}G_l(r, r^{\prime}) = \int \int G(r, \vartheta, \varphi ;
r^{\prime}, \vartheta^{\prime}, \varphi^{\prime}) Y^*_{lm} (\vartheta,
\varphi) Y_{lm} (\vartheta^{\prime}, \varphi^{\prime}) do do^{\prime}
\end{equation}
is the Green's function satisfying
\begin{equation}
\label{G:3} r^2 \left[ - {\frac{ d^2 }{dr^2 }} - {\frac{2 }{r}}
{\frac{d} {dr}} + {\frac{l (l + 1) }{r^2}} + C + i \omega + V(r) \right]
G_l(r, r^{\prime}) = \delta(r - r^{\prime}).
\end{equation}
Eq. (\ref{G:3}) follows from Eq. (\ref{G}) if we first rewrite it in the
spherical coordinates $r, \vartheta, \varphi$, multiply both sides by
$Y^*_{lm}(\vartheta, \varphi) Y_{lm}(\vartheta^{\prime},
\varphi^{\prime})$ and then integrate over $do$ and $do^{\prime}$,
taking into account the hermiticity of the angular part of the Laplacian
and the orthonormality of the spherical harmonics.

Using the above mentioned properties, we can now write Eq.
(\ref{disp:gen}) for the spherically-symmetric AS of radius ${\cal R}_s$
in the form
\begin{equation}
\label{d:3}i \alpha \omega + {\frac{\epsilon^2 l (l + 1) }{{\cal R}_s^2
    }} + \lambda_0 = - \epsilon B Z^{-1} R_{l}(\omega),
\end{equation}
where $\lambda_0$ is the contribution from $\hat{H}_\theta^1$. Eq.
(\ref{d:3}) is the dispersion relation for a fluctuation characterized
by the number $l$.

It can be shown by direct calculation that to the first order in
$\epsilon / {\cal R}_s$ the value of $\lambda_0$ is zero. This means
that in order to calculate $\lambda_0$ from its definition one should
know the distributions $\theta(x)$ and $\eta(x)$ for the AS with greater
accuracy than that of Eq. (\ref{shsm}). Also, the second order of the
perturbation theory has to be taken into account. However, these
difficulties can be avoided, if we use the fact that the system
possesses translational invariance. Indeed, Eq. (\ref{R:3}) should be
identically satisfied for $l = 1$ and $\omega = 0$, so we immediately
obtain that
\begin{equation}
\label{lam0:3}\lambda_0 = - {\frac{2 \epsilon^2 }{{\cal R}_s^2}} -
\epsilon B Z^{-1} R_1(0).
\end{equation}

As we will show below, the instability of a spherically symmetric AS
occurs at ${\cal R}_s \ll 1$. To find the function $R_l(\omega)$
let us use the idea of the previous Section and seek for the Green's
function of Eq. (\ref {G:3}) , treating the potential $V(r)$ as a
perturbation.  The zeroth order Green's function is well known
\begin{equation}
\label{G0:3}G_l^{(0)} (r^{\prime}, r^{\prime\prime}) = \left\{
\begin{array}{ll}
  {\frac{I_{l+1/2}(\kappa r^{\prime}) K_{l+1/2}(\kappa r^{\prime\prime})
      } {\sqrt{r^{\prime}r^{\prime\prime}}}}, & r^{\prime}<
  r^{\prime\prime}, \\ {\frac{I_{l+1/2}(\kappa r^{\prime\prime})
      K_{l+1/2}(\kappa r^{\prime}) }
    {\sqrt{r^{\prime}r^{\prime\prime}}}}, & r^{\prime}>
  r^{\prime\prime},
\end{array}
\right.
\end{equation}
where $\kappa = \sqrt{C + i \omega}$;~ $K_{l+1/2}(z)$ and $I_{l+1/2}(z)$
are the modified Bessel functions. The corrections to
$G_l^{(0)}(r^{\prime}, r^{\prime\prime})$ are then given by
\begin{equation}
\label{Gn:3}G_l^{(n)} (r^{\prime}, r^{\prime\prime}) = - \int_0^\infty
V(r) G_l^{(0)}(r^{\prime}, r) G_l^{(n-1)} (r, r^{\prime\prime}) r^2 dr.
\end{equation}
As a result, from Eq. (\ref{G0:3}) we obtain that
\begin{equation}
\label{Rl:3}R_l^{(0)}(\omega) = {\cal R}_s I_{l+1/2} (\kappa {\cal R}_s) 
K_{l+1/2} (\kappa {\cal R}_s),
\end{equation}
and, as can be seen from Eq. (\ref{Gn:3}), $R_l^{(n)}(\omega) \equiv
{\cal R}_s^2 G_l^{(n)}( {\cal R}_s, {\cal R}_s) = {\cal O}( {\cal
  R}_s^{2n + 1})$.

As before, we will use the values of $B$, $C$, and $Z$ evaluated at the
point $A = A_b$ where ${\cal R}_s \rightarrow 0$ as $\epsilon
\rightarrow 0$.  Expanding the Bessel functions at ${\cal R}_s \ll 1$,
we may write Eq.  (\ref{d:3}) as 
\begin{equation}
\label{d:33}i \alpha \omega + \epsilon^2 {\cal R}_s^{-2} (l + 2) (l - 1)
= \epsilon B Z^{-1} {\cal R}_s \left( {\frac{1 }{3}} - {\frac{1 }{2 l +
    1}} \right).
\end{equation}

Eq. (\ref{d:33}) describes the fluctuations in the case $| \omega |
{\cal R}_s^2 \ll 1$. This is satisfied for the thresholds of the
aperiodic instabilities. Simple calculation shows that an AS becomes
unstable with respect to the aperiodic $l = 0$ mode when
\begin{equation}
\label{Rb:3}{\cal R}_s < {\cal R}_b \equiv \left( {\frac{ 3 \epsilon Z
    }{B}} \right)^{1/3}.
\end{equation}
For ${\cal R}_s > {\cal R}_b$ it becomes unstable with respect to the
aperiodic fluctuations with $l > 1$ at
\begin{equation}
\label{Rcl:3}{\cal R}_s > {\cal R}_{cl} \equiv \left( {\frac{3 (l + 2) (
    2l + 1) \epsilon Z }{2 B }} \right)^{1/3}.
\end{equation}
One can see from this equation that the first instability point
corresponds to $l = 2$:
\begin{equation}
\label{Rc2:3}{\cal R}_{c2} = \left( {\frac{ 30 \epsilon Z }{B}}
\right)^{1/3}.
\end{equation}
Thus, a spherically-symmetric AS can be stable only if its radius
satisfies ${\cal R}_b < {\cal R}_s < {\cal R}_{c2}$, where ${\cal R}_b$
and ${\cal R}_{c2}$ are given by Eqs. (\ref{Rb:3}) and (\ref{Rc2:3}),
respectively.

The $l = 1$ mode corresponds to the translation of the AS as a whole. As
in the case of the one-dimensional AS, for some value of $\alpha \ll 1$
the spherically-symmetric AS becomes unstable with respect to 
the fluctuation leading to the formation of the traveling AS. The
instability with respect to the $l = 1$ mode occurs for some ${\cal R}_s
> {\cal R}_T$ when $\omega \rightarrow 0$. Expanding the Bessel
functions in the right-hand side of Eq. (\ref{d:3}) for small ${\cal
  R}_s$ and $\omega$, we get
\begin{equation}
\label{dt:3}i \alpha \omega = \epsilon B Z^{-1} {\cal R}_s^3 \left\{ i
\omega {\frac{2 }{15}} - (i \omega)^2 {\frac{{\cal R}_s }{24 \sqrt{C}}}
\right\} + {\rm h.~ o.~ terms.}
\end{equation}
According to this equation, the static AS transforms into the traveling
one, when ${\cal R}_s > {\cal R}_T$, where
\begin{equation}
\label{Rt:3}{\cal R}_T = \left( {\frac{ 15 \alpha Z }{2 \epsilon B}}
\right)^{1/3}.
\end{equation}

Now let us study the instabilities with $Re~\omega \not= 0$. Consider an
AS stable when $\alpha \gg 1$, and whose radius is therefore of order
$\epsilon^{1/3}$. As follows from Eq. (\ref{d:3}), in order for an
instability to occur the frequency $Re~\omega$ at the threshold of the
instability should be big enough, so that the argument of the Bessel
functions in Eq. (\ref{Rl:3}) is of order one. This means that $\omega
\sim {\cal R}_s^{-2} \sim \epsilon^{-2/3}$ and therefore the critical
values of $\alpha$ are of order $\epsilon^2$. Indeed, let us introduce
the new variables
\begin{equation}
\label{scale}{\bar{\alpha}} = \alpha / \epsilon^2,~~ \bar{\omega} =
\omega \left( {\frac{\epsilon Z }{B}} \right)^{2/3},~~ p = {\frac{B
    {\cal R}_s^3 }{ \epsilon Z}}.
\end{equation}
Substituting these variables into Eqs. (\ref{d:3}), (\ref{lam0:3}) and
(\ref {Rl:3}), after some algebra we obtain the following transcendent
equation which has explicit dependence on ${\bar{\alpha}}$, $p$, and
$\bar{\omega}$ only.
\begin{equation}
\label{trans:3}i {\bar{\alpha}} \bar{\omega} + (l + 2) (l - 1) p^{-2/3}
= p^{1/3} \left(\frac{1}{3} - I_{l+1/2} \left\{ p^{1/3}
{\bar{\omega}}^{1/2} \sqrt{i} \right\} K_{l+1/2} \left\{ p^{1/3}
{\bar{\omega}}^{1/2} \sqrt{i} \right\} \right).
\end{equation}

We solved Eq. (\ref{trans:3}) numerically for $l = 0$ and 2 in the
region of $p$ where the AS is stable with respect to the aperiodic
fluctuations. The resulting stability diagram is shown in Fig.
(\ref{puls3d}). The rescaled critical frequency as a function of
${\bar{\alpha}}$ for $l = 0$ is also presented in Fig. (\ref{omegac3d}).
From Fig. (\ref{puls3d}) it is clear that when $\alpha$ gets smaller,
the AS always loses stability with respect to the $l = 0$ pulsations
first. The AS is always unstable if $\alpha < \alpha_c = 4.4
\epsilon^2$. For $\alpha > 6.7 \epsilon^2$ the AS destabilizes with
respect to the aperiodic $l = 2$ mode first, if its radius is increased.
All other instabilities, including the one leading to the formation of a
traveling AS, occur at smaller values of $\alpha$ and are, therefore,
secondary.

Let us now consider the case of radially-symmetric AS in two and three
dimensions. Because of the close analogy with the case of
spherically-symmetric AS, we will only outline the derivations, focusing
mainly on the obtained instability criteria.

If $r$, $\varphi$ and $z$ are the cylindrical coordinates, then the
coordinate $\rho = - (r - {\cal R}_s)$, and the operator $\hat{S}$ is
\begin{equation}
\label{S:2}\hat{S} = - \frac{1}{{\cal R}_s^2} {\frac{\partial ^2
    }{\partial \varphi^2}} - {\frac{\partial ^2 }{\partial z^2}}.
\end{equation}
The eigenfunctions of this operator are $\phi_{km} = (4 \pi^2{\cal
  R}_s)^{-1/2} e^{i k z + i m \varphi }$ with the eigenvalues $\nu_{km}
= k^2 + \frac{m^2}{{\cal R}_s^2}$.

The non-vanishing matrix elements of the response operator in this case
are
\begin{equation}
\label{R:2}\langle mk | \hat{R} |mk^{\prime}\rangle \equiv - B R_m(k,
\omega) \delta (k - k^{\prime}),
\end{equation}
where the function $R_m(k, \omega)$ can be expressed in terms of the
Green's function given by
\begin{equation}
\label{G:2}r \left[ - {\frac{d^2 }{d r^2}} - \frac{1}{r} \frac{d}{dr} + 
{\frac{m^2 }{ r^2}} + C + k^2 + i \omega + V(r) \right] G_{km} (r,
r^{\prime}) = \delta (r - r^{\prime}),
\end{equation}
as $R_m (k, \omega) = {\cal R}_s G_{km}({\cal R}_s, {\cal R}_s)$.

The dispersion relation for the fluctuations with particular values of
$k$ and $m$, obtained from Eq. (\ref{disp:gen}), is
\begin{equation}
\label{disp:2}i \alpha \omega + \epsilon^2 k^2 + {\frac{\epsilon^2 m^2 } 
  {{\cal R}_s^2}} + \lambda_0 = - \epsilon B Z^{-1} R_m(k, \omega).
\end{equation}
Because of the translational invariance, this equation should be
satisfied identically for $k = 0$, $\omega = 0$ and $m = 1$. This gives
us
\begin{equation}
\label{lam0:2}\lambda_0 = - {\frac{\epsilon^2 }{{\cal R}_s^2}} -
\epsilon B Z^{-1} R_1(0, 0).
\end{equation}

As before, for small values of ${\cal R}_s$ we will seek for the
function $R_m(k, \omega)$ perturbatively. The zeroth order Green's
function here is
\begin{equation}
\label{G0:2}G_{km}^{(0)} (r^{\prime}, r^{\prime\prime}) = \left\{
\begin{array}{ll}
  I_m (\kappa r^{\prime}) K_m(\kappa r^{\prime\prime}), & r^{\prime}<
  r^{\prime\prime}, \\ I_m (\kappa r^{\prime\prime}) I_m(\kappa
  r^{\prime}), & r^{\prime}> r^{\prime\prime},
\end{array}
\right.
\end{equation}
where $\kappa = \sqrt{C + k^2 + i \omega}$, $I_m(z)$ and $K_m(z)$ are
the modified Bessel functions. The corrections to the Green's function
are given by
\begin{equation}
\label{Gn:2}G_{km}^{(n)} (r^{\prime}, r^{\prime\prime}) = -
\int_0^\infty V(r) G_{km}^{(0)} (r^{\prime}, r) G_{km}^{(n-1)} (r,
r^{\prime\prime}) r dr.
\end{equation}
From this we find that
\begin{equation}
\label{R0:2}R_{m}^{(0)}(k, \omega) = {\cal R}_s I_m( \kappa {\cal R}_s) 
K_m( \kappa {\cal R}_s),
\end{equation}
and that the corrections to $R_m(k, \omega)$ from $G_{km}^{(n)}$ are
$R_m^{(n)} (k, \omega) \equiv {\cal R}_s G_{km}^{(n)} ({\cal R}_s, {\cal
  R}_s) = {\cal O} ({\cal R}_s^{2n + 1})$.

Let us study the stability of the radially-symmetric AS in two
dimensions first. Putting $k = 0$ in Eq. (\ref{disp:2}) with $R_m(k,
\omega)$ given by Eq. (\ref{R0:2}), we obtain that for $m > 1$ and
${\cal R}_s \ll 1$ the aperiodic instability occurs at ${\cal R}_s
> {\cal R}_{cm}$, where
\begin{equation}
\label{Rcn:2}{\cal R}_{cm} = \left( {\frac{2 \epsilon m (m + 1) Z }{B}} 
\right)^{1/3}.
\end{equation}
According to this equation, when the value of ${\cal R}_s$ is increased
the AS becomes unstable with respect to the fluctuation with $m = 2$
when ${\cal R}_s > {\cal R}_{c2}$, where
\begin{equation}
\label{Rc2:2}{\cal R}_{c2} = \left( {\frac{12 \epsilon Z }{B}}
\right)^{1/3}.
\end{equation}

When the value of ${\cal R}_s$ is decreased, at ${\cal R}_s < {\cal
  R}_b$ the radially-symmetric AS in two dimensions becomes unstable
with respect to the aperiodic fluctuations with $m = 0$. According to
Eq. (\ref{disp:2}), for small values of ${\cal R}_s$ we have
\begin{equation}
\label{Rb2:2}{\cal R}_b =
\left( {\frac{ \epsilon Z }{B b_1}} \right)^{1/3},
\end{equation}
where $b_1 = - \log( 1.47 {\cal R}_s \sqrt{C} )$. Since $b_1$ only
weakly depends on ${\cal R}_s$, for ${\cal R}_s \sim 0.1$ we may put
$b_1 \simeq 2$, and for ${\cal R}_s \sim 0.01$ we may put $b_1 \simeq
4$. Thus, a static radially-symmetric AS in two dimensions can be stable
only if ${\cal R}_b < {\cal R}_s < {\cal R}_{c2}$ with ${\cal R}_b$ and
${\cal R}_{c2}$ given in Eqs. (\ref{Rb2:2}) and (\ref{Rc2:2}),
respectively.

When $\alpha$ gets sufficiently small, the radially-symmetric AS in two
dimensions typically destabilizes with respect to the $m = 0$ pulsations
first. If we introduce the variables of Eq. (\ref{scale}) into Eq.
(\ref{disp:2}), for small ${\cal R}_s$ we will obtain an equation
similar to Eq. (\ref{trans:3}), which depends on $p$, $\bar{\alpha}$,
and $\bar{\omega}$ only. Solving this equation numerically for $m = 0$
and 2, (the case $m = 1$ will be discussed below) we obtain the
stability diagram for the radially-symmetric AS in two dimensions (Fig.
\ref{puls2d}). Figure \ref{puls2d} shows that when $\alpha < \alpha_c =
6.1 \epsilon^2$, the AS is unstable for all values of ${\cal R}_s$. When
$\alpha > 11 \epsilon^2$ the AS first destabilizes with respect to the
aperiodic fluctuations with $m = 2$. The dependence of the rescaled
critical frequency $\bar{\omega}$ on ${\bar{\alpha}}$ for the $m = 0$
pulsations is presented in Fig. \ref{omegac2d}. All instabilities with
$m > 2$ occur at smaller values of $\alpha$, so they are secondary.

Now let us turn to the radially-symmetric AS in three dimensions.
Solving the dispersion equation for small ${\cal R}_s$ for $m = 0$, we
obtain that the AS becomes unstable when ${\cal R}_s < {\cal R}_b$,
where
\begin{equation}
\label{Rb:2}{\cal R}_b = \left( {\frac{ 2.25 \epsilon Z }{B}}
\right)^{1/3},
\end{equation}
or when ${\cal R}_s > {\cal R}_c$, where
\begin{equation}
\label{Rc:2}{\cal R}_c = \left( {\frac{ 7.5 \epsilon Z }{B}}
\right)^{1/3},
\end{equation}
when
\begin{equation}
\label{kc:2}k = k_c = \left( {\frac{4.1 \epsilon Z }{B}} \right)^{-1/3}. 
\end{equation}
Comparing Eq. (\ref{Rc:2}) with (\ref{Rc2:2}), one can see that in
contrast to the radially-symmetric AS in two dimensions, when ${\cal
  R}_s$ is increased, the radially-symmetric AS in three dimensions
destabilizes with respect to $m = 0$ mode first.

In the case $m = 1$ an AS destabilizes with respect to the fluctuations
with small $k$ at $Re~\omega = 0$. Expanding the Bessel functions in the
right-hand side of Eq. (\ref{disp:2}), we will obtain the following
equation
\begin{equation}
\label{anti:2}i \alpha \omega + \epsilon^2 k^2 = \epsilon B Z^{-1} b_2
{\cal R}_s^3 (k^2 + i \omega),
\end{equation}
where $b_2 = - \log (1.14 {\cal R}_s \sqrt{C} ) / 4$. The value of $b_2$
weakly depends on ${\cal R}_s$, so for ${\cal R}_s \sim 0.1$ we may put
$b_2 \simeq 0.5$, and for ${\cal R}_s \sim 0.01$ we may put $b_2 \simeq
1$. It can be seen from Eq. (\ref{anti:2}) that when $\alpha \gg
\epsilon^2$ an AS becomes unstable when ${\cal R}_s > {\cal R}_{c1}$ at
$k \rightarrow 0$, where
\begin{equation}
\label{rc1:2}{\cal R}_{c1} = \left( {\frac{\epsilon Z }{B b_2}}
\right)^{1/3}.
\end{equation}
Comparing Eq. (\ref{rc1:2}) with Eq. (\ref{Rb:2}) one can see that for
$\epsilon \lesssim 10^{-3}$ we have ${\cal R}_{c1} < {\cal R}_b$, and,
therefore, the cylindrically-symmetric AS is always unstable. However,
note that the increment of the fluctuations with $m = 1$ and small $k$
may be extremely small.

Similarly, as follows from Eq. (\ref{anti:2}), when $\alpha \sim
\epsilon^2$ the AS becomes unstable for ${\cal R}_s > {\cal R}_T$ at $k
= 0$, where
\begin{equation}
\label{Rt:2}{\cal R}_T = \left( {\frac{ \alpha Z }{\epsilon B b_2}}
\right)^{1/3},
\end{equation}
and transforms into the traveling AS. As we already noticed, when ${\cal
  R}_s$ is not very small, the instability leading to the formation of a
traveling AS occurs at smaller values of $\alpha$ than the instability
with respect to the $m = 0$ pulsations. However, when ${\cal R}_s
\lesssim 0.01$ the coefficient $b_2$ in Eq. (\ref{Rt:2}) becomes such
that the line $p(\bar{\alpha})$ in Fig. (\ref{puls2d}) corresponding to
the instability with $m = 1$ crosses the curve corresponding to the
threshold of the $m = 0$ pulsations when $p < 12$. In this situation a
radially-symmetric AS in two or three dimensions may become unstable and
transform into the traveling AS before it destabilizes with respect to
the $m = 0$ pulsations. However, this is a rather unrealistic situation
since in order for an AS to have a radius ${\cal R}_s \lesssim 0.01$ and
be unstable with respect to the $m = 1$ mode, one should have $\epsilon
\lesssim 10^{-6}$ and $\alpha \lesssim 10^{-12}$.

\section{conclusion}

Thus, in the present paper we developed an asymptotic theory of the
instabilities of arbitrary $d$-dimensional static patterns which may
form in a wide class of reaction-diffusion systems of the
activator-inhibitor type. This theory is based on the natural smallness
of the parameter $\epsilon = l / L$, where $l$ and $L$ are the
characteristic length scales of the activator $\theta$ and the inhibitor
$\eta$, respectively. In fact, as was already mentioned, if the length
scales $l$ and $L$, as well as the normalization of $\theta$ and $\eta$,
are chosen properly, the condition $\epsilon \ll 1$ is not only
sufficient, but also necessary for the existence of AS and other large
amplitude patterns \cite{ko:book,ko:ufn89}.

Within the presented theory we analyzed different types of spontaneous
transformations of the simplest static patterns in two- and
three-dimensional systems into more complex static, pulsating, and
traveling patterns. We showed that the criteria corresponding to these
transformations are universal in the sense that they are practically
independent of the specific nonlinearities of the system and are
determined only by the two parameters: $\epsilon$ and $\alpha$; and
three numerical constants: $B$, $C$, and $Z$, which have all necessary
information about the nonlinearities. If the length scales and the
normalization of $\theta$ and $\eta$ are chosen properly, the constants
$B$, $C$, and $Z$ are necessarily of order one, and the constant $B$ can
in principle be small.

Let us summarize the results of our analysis.

According to the formulas of Sec. IV, when $\alpha \gg \epsilon$ a
one-dimensional AS in two and three dimensions (stripe) is less stable
than the AS in one dimension. As follows from Eqs. (\ref{corr}) and
(\ref{Lc2}), for $\epsilon \ll 1$ we have ${\cal L}_{c2} < {\cal
  L}_{c1}$. This means that when the control parameter $A$ is increased
and the AS widens, the AS always destabilizes with respect to the
wriggling of its walls first.  It is natural to expect that for ${\cal
  L}_{c2} < {\cal L}_s < {\cal L}_{c1}$ a one-dimensional AS will deform
into a twisted band that fills the whole volume of the system. When $A$
is further increased so that ${\cal L}_s > {\cal L}_{c1}$, each wall of
the pattern becomes unstable with respect to the fluctuations with the
characteristic wave vectors $k_c \sim \epsilon^{-1/3}$.  As a result of
the development of this instability, fingers will start to grow from the
walls of the pattern. Eventually, the volume of the system will become
filled with a labyrinthine pattern. The instability will persist until
the distance between the walls and their curvature radii become of order
$\epsilon^{1/3}$. This phenomenon was observed recently in the
experiments by Lee and Swinney \cite{lee:pre95} and in the numerical
simulations of a two-dimensional reaction-diffusion system
\cite{hagberg:prl94,petrich94}.

When $A$ is decreased and the AS narrows, at ${\cal L}_s < {\cal
  L}_{cb}$ it destabilizes with respect to the corrugation of its walls
with the wavevector $k_c \sim \epsilon^{-1/3}$. As a result, granulation
of the static AS will occur and eventually the resulting granules with
the radius of order $\epsilon$ will disappear.

According to Eqs. (\ref{Lc2}) and (\ref{Lt}), for $\alpha < \epsilon^2$
we have ${\cal L}_T < {\cal L}_{c2}$. This means that as ${\cal L}_s$ is
increased, a one-dimensional AS will always transform to a traveling
band first. Note that the condition $\alpha < \epsilon^2$ here is exact
and does not depend in any way on the nonlinearities of the system.

As follows from Eqs. (\ref{puls}) and (\ref{Lt}), there may be a rather
wide range of the parameters $\alpha$ and $\epsilon$ for which the
pulsation instability emerges before the instability leading to the
formation of traveling AS as the width of the AS is increased or the
value of $\alpha$ is decreased. Indeed, the condition ${\cal L}_\omega <
{\cal L}_T$ is satisfied when
\begin{equation} \label{pt} \alpha > \epsilon \beta_\omega, 
\end{equation} 
where
\begin{equation} \label{beta} \beta_\omega = { B \over
    650 Z C^{3/2}}.
\end{equation} 
Note the huge numerical factor in the denominator of Eq. (\ref{beta}).
Because of it the instability with respect to pulsations will emerge
before the instability to traveling AS in the majority of the real
systems. At the same time, as follows from Eqs.  (\ref{puls}) and
(\ref{Lc2}), the instability with respect to pulsations is the first, i.
e., ${\cal L}_\omega < {\cal L}_{c2}$, if
\begin{equation} \label{pc} 
  \alpha < \epsilon^{5/2} \beta_\omega^{-1/2}.
\end{equation} 
These two conditions can be satisfied at the same time only if
\begin{equation} \label{e} 
  \epsilon > \beta_\omega.
\end{equation} As a result of the instability with
respect to pulsations a static AS may transform into a stationary
breathing pattern (pulsating AS), if the parameters of the system are
finely adjusted \cite{ko:book,ko:ufn89}. However, as we see from our
numerical simulations, in most cases the walls of the AS go so far apart
that a local breakdown occurs at the AS center, and eventually the AS
produces two one-dimensional AS traveling in the opposite directions.

When the width of the AS is decreased, at $\alpha < \epsilon^2$ the AS
destabilizes with respect to pulsations and disappears, if ${\cal L}_s <
{\cal L}_{b\omega}$, where ${\cal L}_{b\omega}$ is given in Eq.
(\ref{Lbom}). When the value of $\alpha$ decreases, at $\alpha =
\alpha_c$ we have ${\cal L}_{b\omega} = {\cal L}_T$.  For smaller values
of $\alpha$ the AS is always unstable. According to Eqs.  (\ref{Lt}) and
(\ref{Lbom}), the value of $\alpha_c$ is given by
\begin{equation} \label{alc}
  \alpha_c \simeq \epsilon^3 \left( \log \epsilon^{-1} \right)^2.
\end{equation}

\begin{figure*}[htb]
\begin{center}
\caption{} Formation of a labyrinthine pattern. Numerical solution of
Eqs. (\ref{act}) and (\ref{inh}) with $q = \theta^3 - \theta - \eta$ and
$Q = \theta + \eta - A$ with $\epsilon = 0.05$, $\alpha = 0.1$, $A =
-0.3$ at different times. The system size is $20 \times 20$. At $t = 0$
the homogeneous state of the system is excited in the region $0.5 \times
0.7$.  Distributions of the activator at times $t = 0, 16.5, 30, 65,
100, 300$.  
\label{fsim} 
\end{center} 
\end{figure*} 
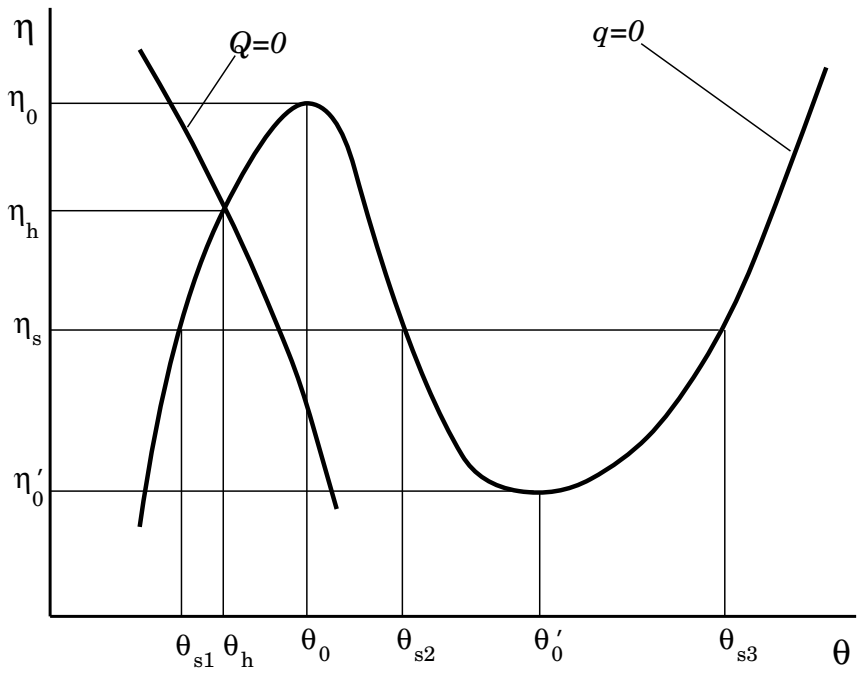
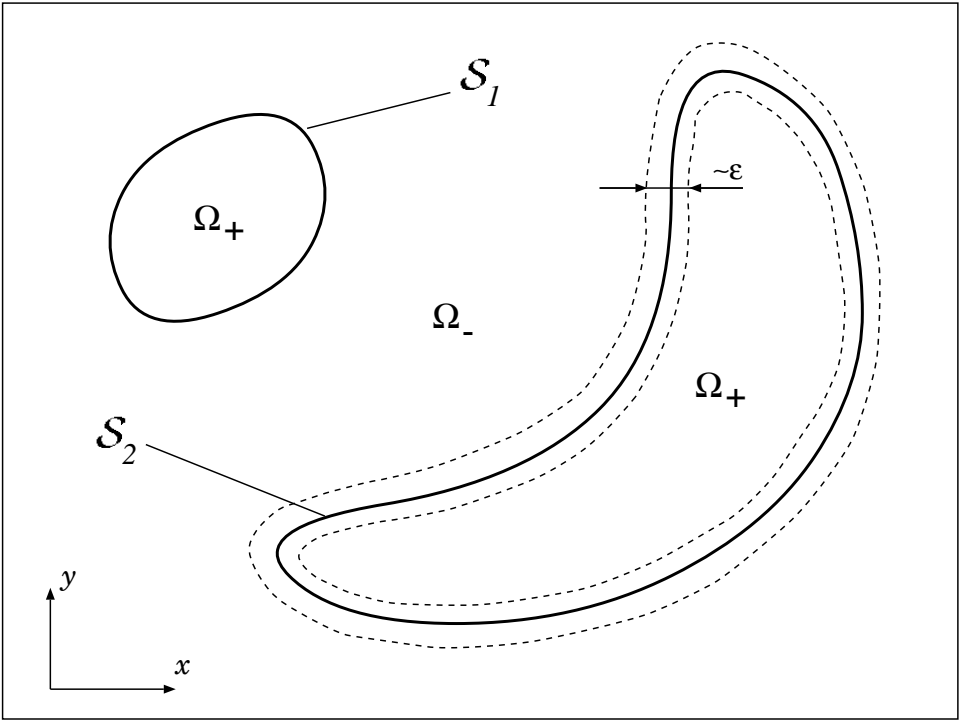
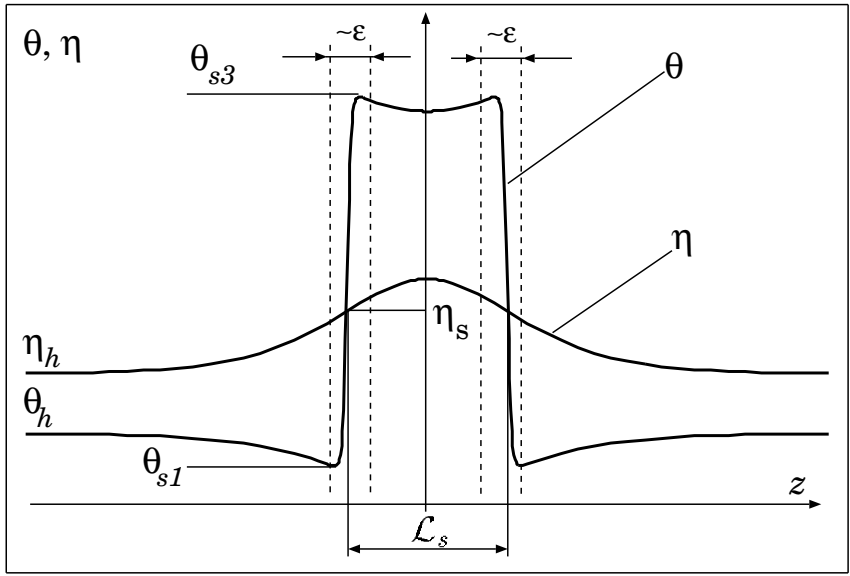
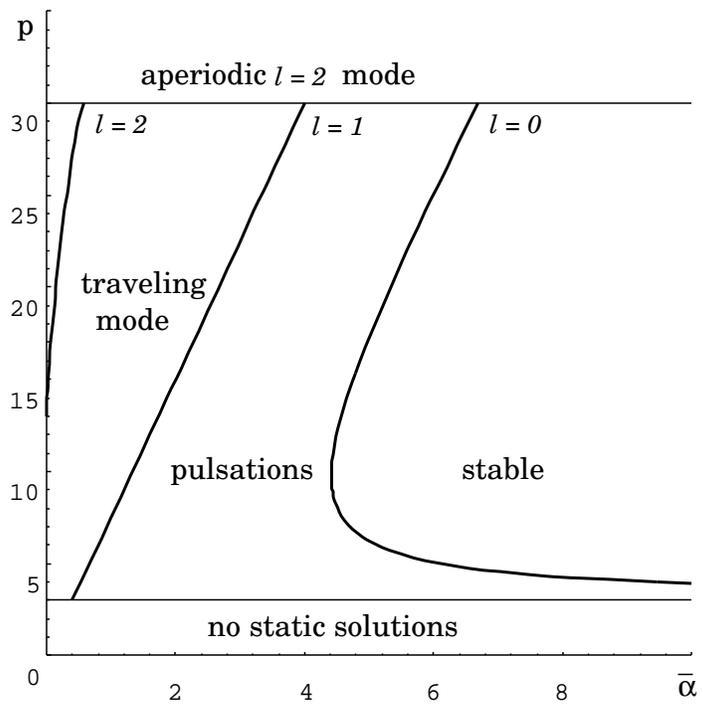
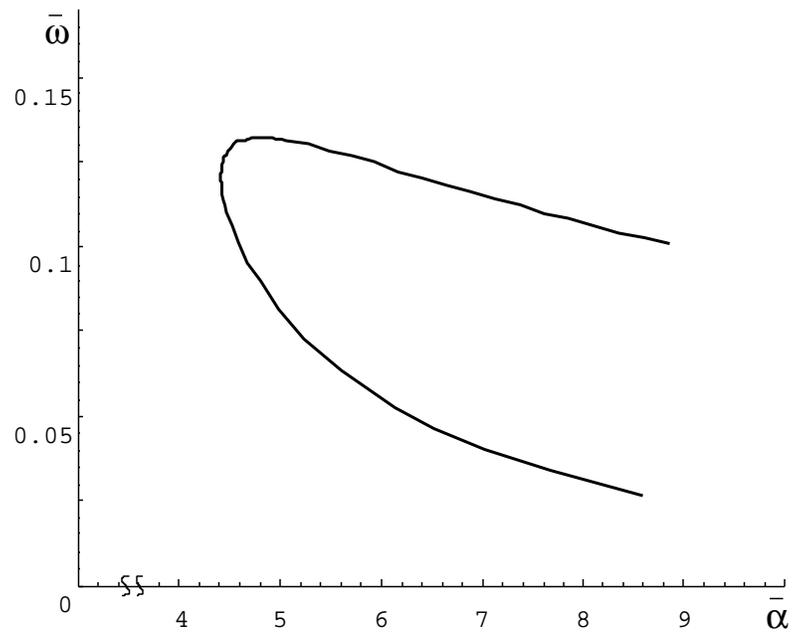
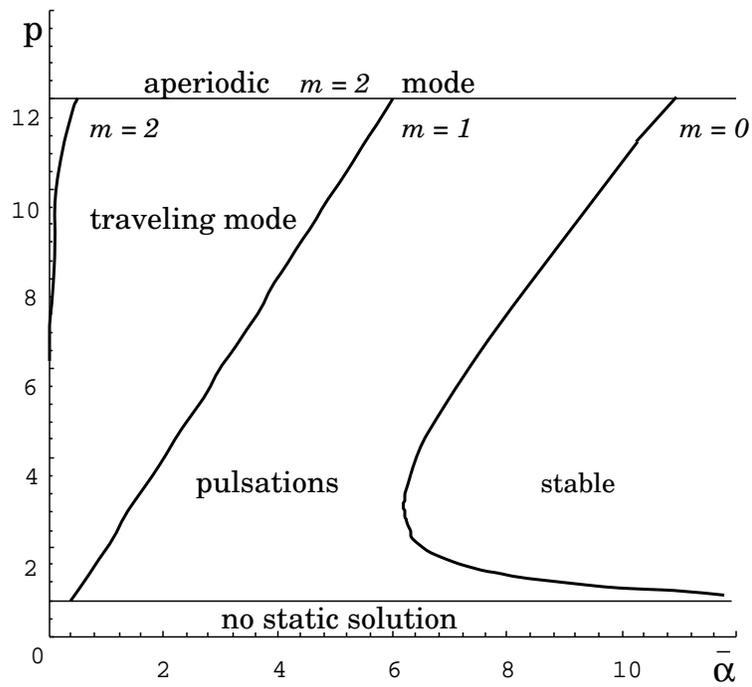
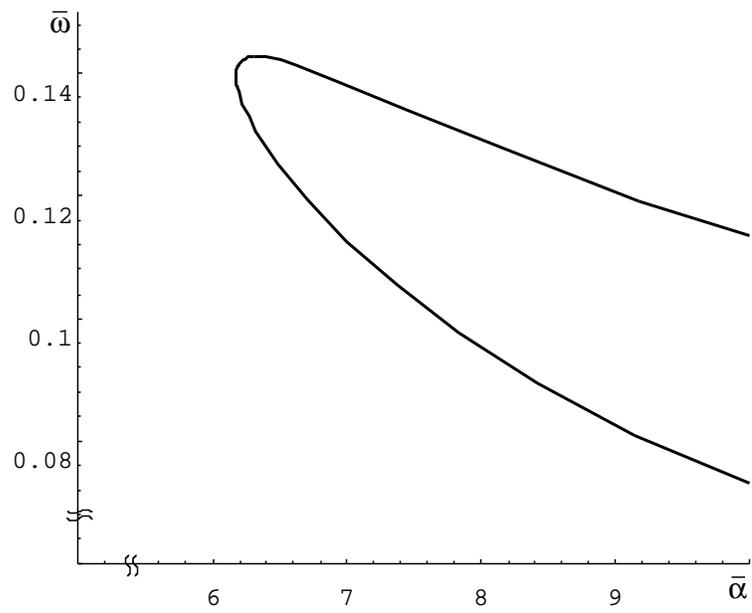
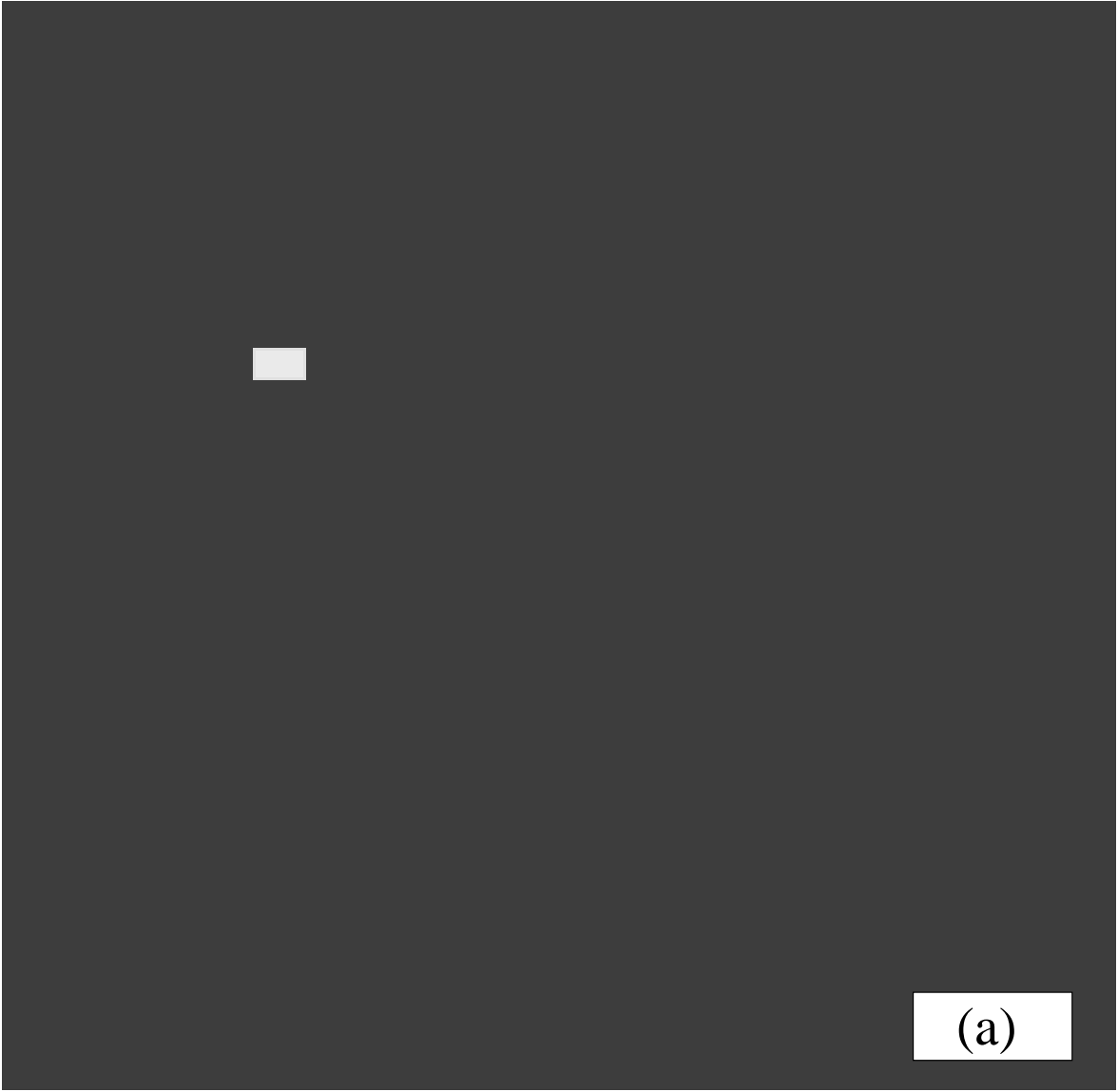
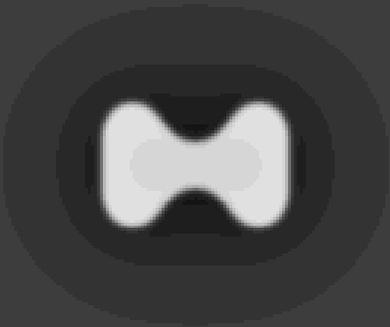
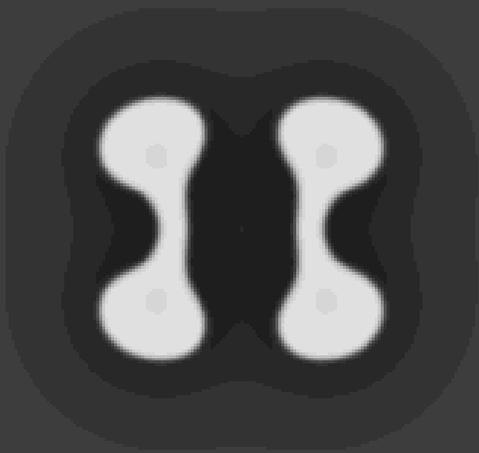
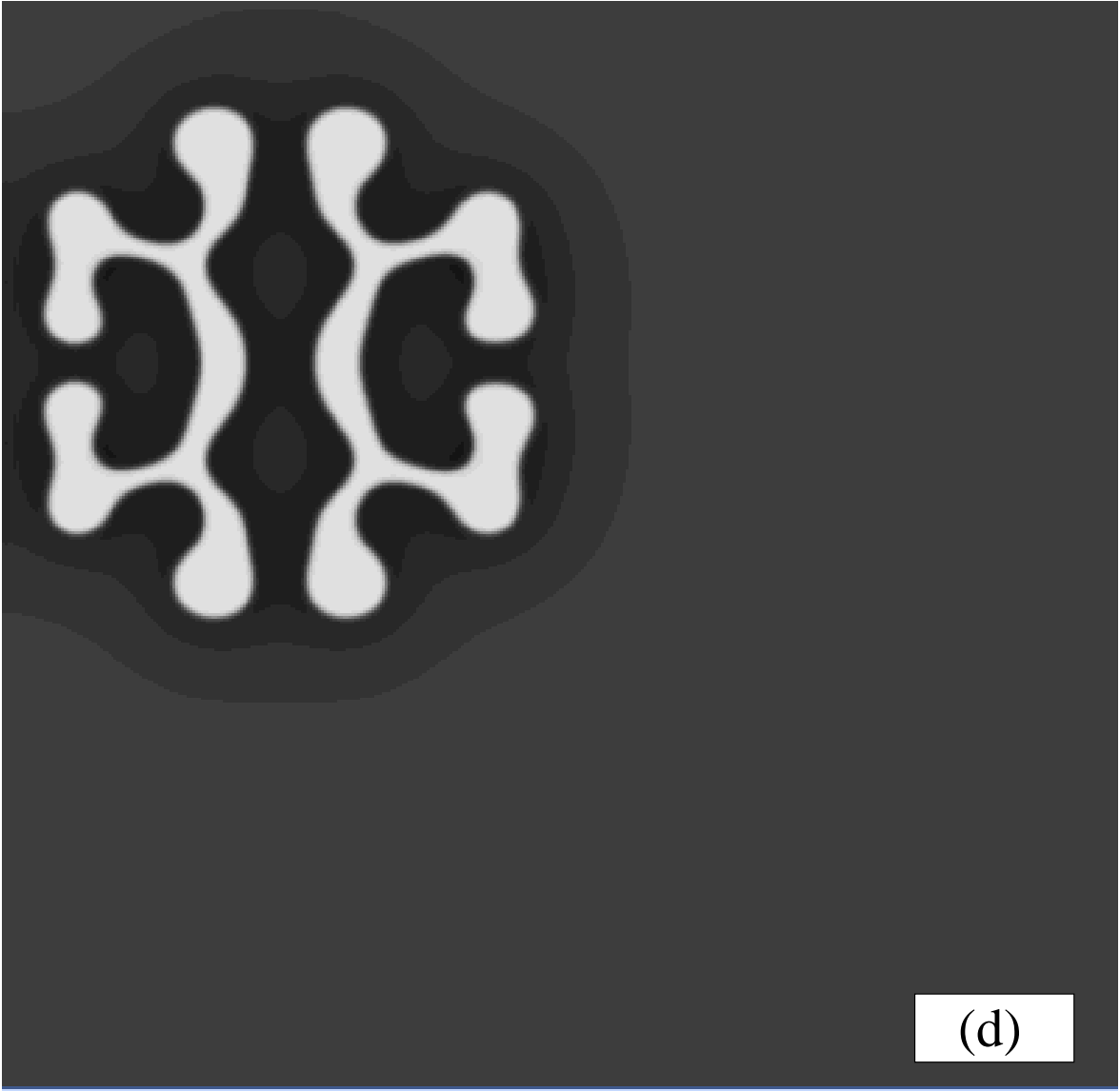
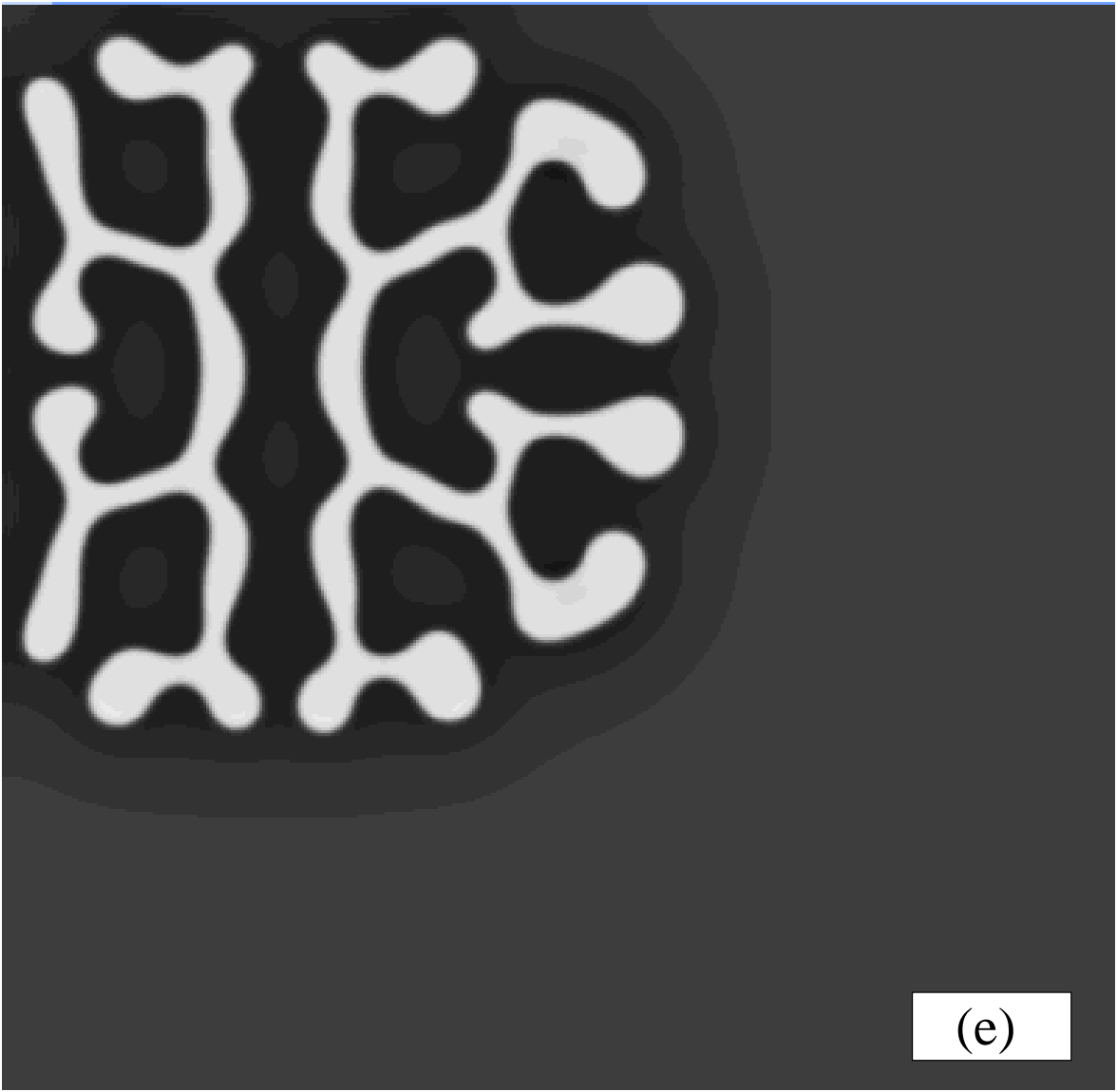
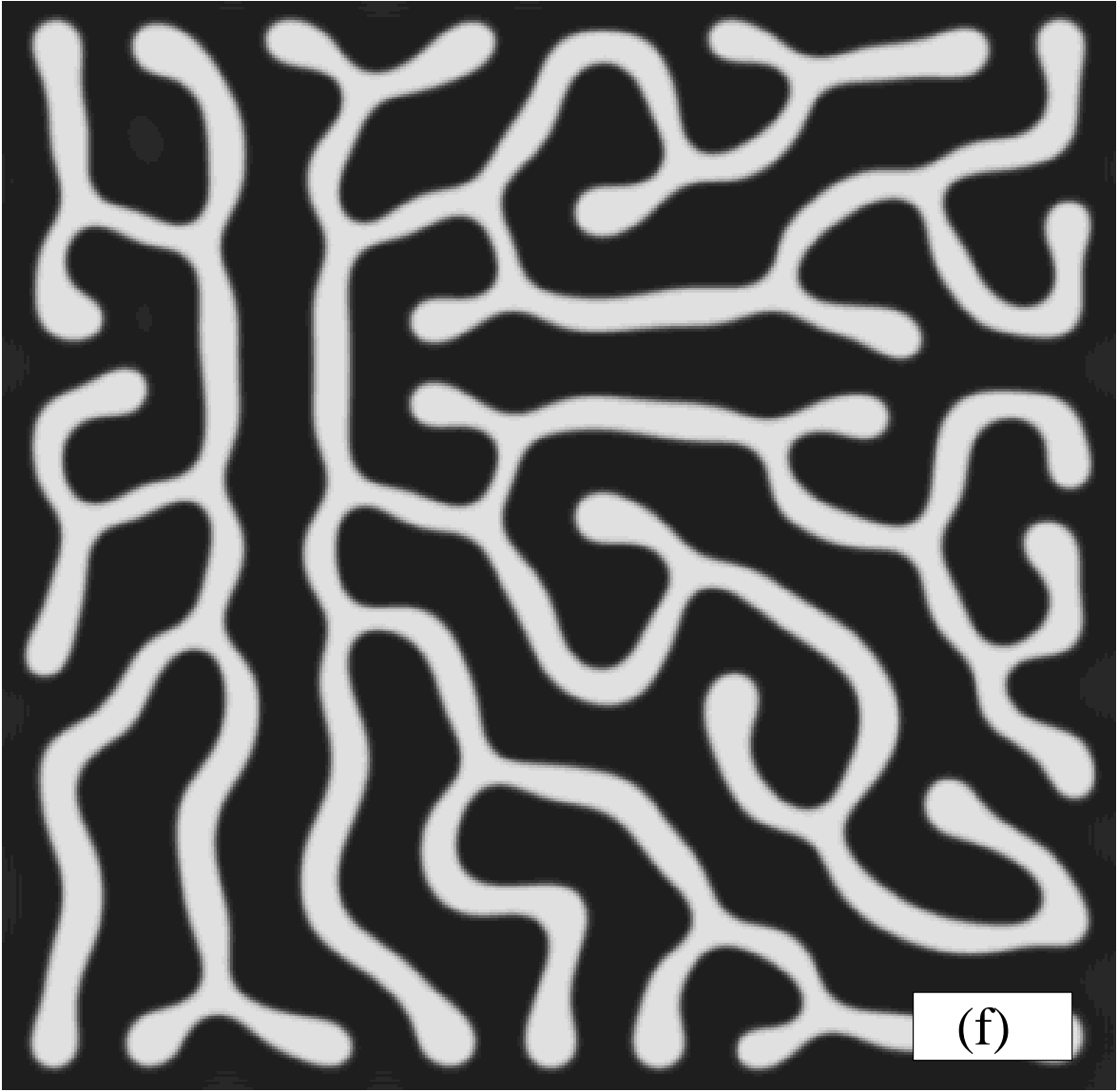

The transformations and the evolution of the static
spherically-symmetric AS in three dimensions and static
radially-symmetric AS in two dimensions are very similar. These AS are
stable only in the relatively narrow range of their radii. When $\alpha$
is big enough, the AS is stable if ${\cal R}_{b} < {\cal R}_s < {\cal
  R}_{c2}$, where ${\cal R}_b$ and ${\cal R}_{c2}$ are of order
$\epsilon^{1/3}$ and are given by Eqs.  (\ref{Rb:3}) and (\ref{Rc2:3})
for the spherically-symmetric, and by Eqs. (\ref{Rb:2}) and
(\ref{Rc2:2}) for the radially-symmetric AS in two dimensions,
respectively. As the control parameter $A$ is decreased, at ${\cal R}_s
< {\cal R}_b$ the AS will abruptly disappear. As $A$ is increased and
the AS radius becomes greater than ${\cal R}_{c2}$, the AS loses
stability with respect to the radially-nonsymmetric fluctuations with $l
= 2$ first. The growth of these fluctuations may lead to the splitting
of the AS into two, or to the growth of a pattern with the sophisticated
geometry. This complex pattern may further get more and more complicated
as a result of the fingering instability, if the distance between the
pattern's walls exceed the value of order $\epsilon^{1/3}$. The process
of splitting and complicating will go on until the whole system becomes
filled with a very sophisticated labyrinthine pattern (Fig. \ref{fsim}).
The pattern should not necessarily be connected because of the
possibility of splitting. Thus, there is a remarkable phenomenon
characteristic to the considered class of nonlinear systems: {\em as a
  result of the instability of the AS localized in a small portion of an
  extended system the whole system becomes filled with a complicated
  pattern}.  These conclusions explain the effects of splitting,
self-replication, and formation of labyrinthine patterns found recently
in the experimental and numerical investigations of some two-dimensional
reaction-diffusion systems \cite{lee:pre95,hagberg:prl94,petrich94}.

Static cylindrically-symmetric AS in three dimensions can only be stable
if $\epsilon \gtrsim 10^{-3}$. If this condition is satisfied, the
cylindrically-symmetric AS will destabilize with respect to wriggling ($
m = 1$ mode) when ${\cal R}_s > {\cal R}_{c1}$, where ${\cal R}_{c1}$ is
given by Eq. (\ref{rc1:2}) as its radius is increased. If the radius of
the AS is decreased, at ${\cal R}_s < {\cal R}_{b}$, where ${\cal R}_b$
is given by Eq. (\ref{Rb:2}), the AS will destabilize with respect to
the corrugation of its walls ($m = 0$ and $k = k_c$, where $k_c$ is
given by Eq. (\ref{kc:2})). As a result, the AS will granulate and
transform into a number of spherically-symmetric AS.

When the value of $\alpha$ is decreased, a stable radially-symmetric AS
in two dimensions and spherically-symmetric AS in three dimensions lose
their stability with respect to the radially-symmetric fluctuations
oscillating with some characteristic frequency (see Figs. \ref{puls3d}
-- \ref{omegac2d}). Only the radially-symmetric AS in two dimensions
whose radius ${\cal R}_s \lesssim 0.01$ can spontaneously transform to a
traveling AS before it destabilizes with respect to the $m = 0$
pulsations. However, as we already mentioned, this situation is possible
only for unrealistically small values of $\alpha$ and $\epsilon$,
therefore, this bifurcation, which was recently discussed in Ref.
\cite{krischer94} is secondary in most of the real reaction-diffusion
systems. As a result of the instability with respect to the
radially-symmetric pulsations the AS may disappear, or, if the
parameters of the system are finely adjusted, a stationary pulsating
radially-symmetric AS may form \cite{ko:book,ko:ufn89}. However, as we
see from our numerical simulations, in most cases the growth of the
amplitude of the AS pulsations leads to the local breakdown in the AS
center and the formation of a traveling wave in the form of a ring with
the radius monotonely growing with time.

Static radially-symmetric AS of any radius are always unstable if
$\alpha < 4.4\epsilon^2$ in three dimensions, or if $\alpha < 6.1
\epsilon^2$ in two dimensions (in the case of extremely small $\epsilon$
and $\alpha$ this value may be greater: see the discussion above). In
this situation only traveling waves and pulsating patterns will form in
the system. Note that these conclusions are totally independent of the
specific nonlinearities of the system.

In our analysis we considered only monostable systems. However, the
results obtained by us remain true in bistable systems as well. In
particular, it can be easily seen that a static one-dimensional front
connecting two stable homogeneous states is always unstable with respect
to the fluctuations with the wavevector $k \sim \epsilon^{-1/3}$.

So, we see that our results are universal and applicable to the wide
class of physical, chemical, and biological systems which can be
described by Eqs. (\ref{1}) and (\ref{2}), if the nullcline of Eq.
(\ref{1}) is N- or inverted N-like (Fig. \ref{f1}). In such N-systems
the universality of the obtained results is related to both the form of
the nullcline and the smallness of the parameter $\epsilon$. At the same
time there are systems which are described by Eqs. (\ref{1}) and
(\ref{2}), in which the nullcline of Eq. (\ref{1}) is V- or
$\Lambda$-like. For example, the biological morphogenesis system of
Gierer and Meinhardt \cite{gierer}, the models of axiomatic chemical
reactions --- the Brusselator \cite{nicolis} and the Gray-Scott model
\cite{pearson:sci93} all have V- or $\Lambda$-like nullclines. In such
V-systems at $\epsilon \ll 1$ spike patterns of giant amplitude may
form \cite{ko:book,ko:ufn89}. The properties of such patterns
essentially differ from those forming in the N-systems we considered
here. For this reason it would be incorrect to use the results of the
present paper to interpret the results of the numerical simulations in
V-systems.  However, the simulations performed by Pearson in the
two-dimensional Gray-Scott model showed \cite{pearson:sci93} that the
effects of the granulation of static one-dimensional AS, splitting and
self-replication leading to the formation of complex patterns which fill
the whole space are seen in V-systems as well.

\appendix

\section{the eigenvalue problem}

\label{eigen}

Let us consider the exact stability problem for a static pattern.
Equations (\ref{tht:xy}) and (\ref{ett:xy}) can be written in operator
form as
\begin{eqnarray}
  \hat{H}_1 \delta \theta = \hat{U}_1 \delta \eta, \label{H_1} \\ 
  \hat{H}_2 \delta \eta = \hat{U}_2 \delta \theta, \label{H_2}
\end{eqnarray}
where
\begin{eqnarray}
  \hat{H}_1 = i \alpha \omega - \epsilon^2 \Delta + q'_\theta,
  \label{HH} \\ \hat{H}_2 = i \omega - \Delta + Q'_\eta,
\end{eqnarray}
and
\begin{equation}
\label{UU}\hat{U}_1 = -q^{\prime}_\eta, ~~~\hat{U}_2 =
-Q^{\prime}_\theta.
\end{equation}
Substituting Eq. (\ref{H_2}) into Eq. (\ref{H_1}), we obtain
\begin{equation}
\label{uhu} ( \hat{H}_1 - \hat{U}_1 \hat{H}_2^{-1} \hat{U}_2 ) \delta
\theta = \lambda \delta \theta,
\end{equation}
where $\lambda$ should be put to zero. Solving this eigenvalue problem
and then requiring that $\lambda = 0$, we may obtain the value of
$\omega$.  In fact, this allows to think of $\lambda$ as an
infinitesimally small quantity.

In the considered problem the operator $\hat{U}_1 \hat{H}_2^{-1}
\hat{U}_2$ may be treated as a perturbation to the operator $\hat{H}_1$
\cite{ko:book,ko:ufn89}.  We would like to find the solution of Eq.
(\ref{uhu}) corresponding to the lowest eigenvalue. In view of the
discussion in Sec. III, to the zeroth order in $\epsilon$ the
eigenfunctions of the operator $\hat{H}_1$ are linear combinations of
the functions $\delta\theta_{il}^{(0)} = \sqrt{\epsilon / Z} ~
\delta\theta_{il}$ where $\delta\theta_{il}$ are defined in Eq.
(\ref{h:w:ef}) and $Z$ is defined in Eq. (\ref{Z}) (the coefficient in
front of $\delta \theta_{il}$ ensures the proper normalization), and
their corresponding eigenvalues are of order $\epsilon$.  It can be
easily seen that
\begin{equation}
\label{00}\langle i^{\prime}l^{\prime}|\hat{U}_1 \hat{H}_2^{-1}
\hat{U}_2 | i l \rangle \sim \epsilon,
\end{equation}
where the matrix element is calculated with the functions
$\delta\theta_{il}^{(0)}$. However, one should be careful in calculating
$\lambda$ since the matrix element from the bound state $\delta
\theta_{il}^{(0)}$ to the state of the continuous spectrum of the
operator $\hat{H}_1$ with the wavevector $k \sim 1$ has the following
estimate
\begin{equation}
\label{0k}\langle k | \hat{U}_1 \hat{H}_2^{-1} \hat{U}_2 | il \rangle
\sim \sqrt{\epsilon}.
\end{equation}
So, the second and higher order corrections of the perturbation theory
given by the transitions from the bound states to the long-wave
continuous spectrum will be of the same order as the first-order
contribution from the diagonal element.

According to Eq. (\ref{uhu}) and the fact that the unperturbed
eigenvalues of the operator $\hat{H}_1$ are of order $\epsilon$, the
improved function $\delta\theta_{il}^{(1)}$ which contains the
corrections of order $\sqrt{\epsilon}$ can be written as
\begin{equation}
\label{1sm}\delta\theta_{il}^{(1)} = \{ 1 + \hat{H}_1^{-1} \hat{U}_1
\hat{H}_2^{-1} \hat{U}_2 + (\hat{H}_1^{-1} \hat{U}_1 \hat{H}_2^{-1}
\hat{U}_2 )^2 + \ldots \} \delta\theta_{il}^{(0)}.
\end{equation}
Of course, as should be in the perturbation theory, the operator
$\hat{H}_1^{-1}$ actually projects out the $\delta \theta_{il}^{(0)}$
components.  If we now substitute this function for $\delta\theta$ into
Eq.  (\ref{uhu}), multiply it by $\delta\theta_{il}^{(0)*}$, and then
integrate over the volume of the system, to the first order in
$\epsilon$ we will arrive at the following equation
\begin{equation}
\label{oper}\langle i^{\prime}l^{\prime}| \hat{H}_1 | il \rangle -
\langle i^{\prime}l^{\prime}| \hat{R} | i l \rangle = \lambda
\delta_{ii^{\prime}} \delta_{ll^{\prime}},
\end{equation}
where
\begin{equation}
\label{RRR}\hat{R} = \hat{U}_1 \hat{H}_2^{-1} \hat{U}_2 \{ 1 +
\hat{H}_1^{-1} \hat{U}_1 \hat{H}_2^{-1} \hat{U}_2 + (\hat{H}_1^{-1}
\hat{U}_1 \hat{H}_2^{-1} \hat{U}_2)^2 + \ldots \}.
\end{equation}
One should not confuse the matrix elements in Eq. (\ref{oper}) with
those of Eq. (\ref{R:me}) and (\ref{disp:gen}), since they are
calculated with the eigenfunctions which have a different normalization.

To the first order in $\epsilon$, we may replace the true distributions
of the activator and the inhibitor in the operator $\hat{R}$ by the
smooth distributions. According to Eq. (\ref{1sm}), the function
$\delta\theta_{sm} = \delta\theta_{il}^{(1)} - \delta\theta_{il}^{(0)}$
has the characteristic length scale 1.  If we neglect the term $\alpha
\omega$ in Eq. (\ref{HH}), the operator $\hat{H}_1^{-1}$ reduces to
$q^{\prime}_\theta(\theta_{sm} (x), \eta_{sm}(x))^{-1}$. Then the
definition of the operator $\hat{R}$ in Eq. (\ref{RRR}), together with
the above mentioned property of the operator $\hat{H}_1$, means that in
the calculation of the inhibitor response one should consider the
fluctuations $\delta\eta$ and $\delta\theta_{sm}$ to be related by Eq.
(\ref {lc:xy}). With all these approximations, Eq. (\ref{oper}) is
equivalent to Eq. (\ref{disp:gen}).

\section{piecewise-linear model}

\label{p-l}

It seems that the only model in which it is possible to find the exact
Green's function of Eq. (\ref{G}) is the well-known piecewise-linear
model of a reaction-diffusions system, which is described by the
following equations \cite{koga80}
\begin{equation} \label{ps:th} 
  \alpha {\partial \theta \over \partial t} = \epsilon^2 \Delta \theta -
  \theta - \eta + H( \theta - A),
\end{equation}
\begin{equation} \label{ps:et} {\partial \eta \over \partial t} =  
  \Delta \eta + \theta - \gamma \eta,
\end{equation} where $H(x)$ is the
Heaviside function.

The homogeneous state of this system is $\theta_h = 0$, $\eta_h = 0$.
It can be easily verified that the values of the parameters describing
the sharp distribution of the activator are $\theta_{s1} = A -
\frac{1}{2}$, $\theta_{s2} = A$, $\theta_{s3} = A + \frac{1}{2}$, and
$\eta_s = \frac{1}{2} - A$
\cite{koga80,kko:mk84:1,kko:mk84:2,kko:dan84}. In view of Eqs.
(\ref{ps:th}), (\ref{ps:et}), (\ref{B}), and (\ref{C}), we obtain that
in this model $B = 1$ and $C = 1 + \gamma$.

In order to find $Z$ we need to know the sharp distribution. According
to Eq. (\ref{sh:1}), for this model
\begin{equation} \label{ps:sh}
  \theta_{sh}(\rho) = \left\{
\begin{array}{ll}
  A - \frac{1}{2} + \frac{1}{2_{ }} \exp( \rho / \epsilon), & \rho < 0,
  \\ A + \frac{1}{2} - \frac{1}{2} \exp( - \rho / \epsilon), & \rho > 0.
\end{array}
\right.
\end{equation} 
From Eqs. (\ref{ps:sh}) and (\ref{Z}) we obtain that $Z = 1/4$. Note
that $B$, $C$, and $Z$ are just constants independent of $A$.

Having calculated the constants $B$, $C$, and $Z$, we can easily find
all instability points. Let us consider the one-dimensional AS, for
example. According to Eq. (\ref{corr}), for $\alpha \gg \epsilon$ the AS
becomes unstable with respect to the symmetric fluctuations with
\begin{equation} \label{ps:kc} 
  k_c = 1.12 \epsilon^{-1/3},~~{\rm at}~~{\cal L}_c = 1.66
  \epsilon^{1/3},
\end{equation} whereas, according to Eq.
(\ref{puls}), in the case $\alpha \ll \epsilon$ it destabilizes
with respect to the fluctuations with 
\begin{equation} \label{ps:oc}
\omega_c = 1.84 \left( {\alpha \over \epsilon} \right)^{-2/3},~~{\rm at
  }~ {\cal L}_\omega = 0.60 \left( {\alpha \over \epsilon}\right)^{1/3}.
\end{equation} 
Eq. (\ref{ps:oc}) improves the accuracy of the results
obtained in Ref. \cite{kuo:pre95}.

Similarly, from Eq. (\ref{b}) one concludes that in this system $b =
1/(2 \sqrt{1 + \gamma})$. According to Eq. (\ref{Lc2}), in the case
$\alpha \gg \epsilon$ a one-dimensional AS becomes unstable with respect
to the antisymmetric fluctuations with $k \rightarrow 0$ at
\begin{equation}
\label{ps:Lc2} {\cal L}_s > {\cal L}_{c2} = (1 + \gamma)^{1/4} (2
\epsilon)^{1/2},
\end{equation} 
what agrees with the result of Ref. \cite{ohta89}. In the case $\alpha
\ll \epsilon$, according to Eq. (\ref{Lt}), a static AS
spontaneously transforms into traveling when
\begin{equation}
\label{ps:Lt} 
{\cal L}_s > {\cal L}_T = (1 + \gamma)^{1/4} \left({ 2 \alpha \over
  \epsilon} \right)^{1/2}.  \end{equation} This result coincides with
the one obtained in Ref. \cite{kuo:pre95}.

When ${\cal L}_s$ becomes comparable with $\epsilon$, an AS becomes
unstable with respect to the symmetric fluctuations. According to Eq.
(\ref{Lb}), for $\alpha \gg \epsilon$ the instability occurs at ${\cal
  L}_s < {\cal L}_b$ with respect to the fluctuations with $k = k_c$,
where 
\begin{equation}
\label{ps:Lb} 
k_c = 1.26 \epsilon^{-1/3}, ~~~~{\cal L}_b = \frac{4 \epsilon}{3} \log
\epsilon^{-1},  
\end{equation} 
and, for $\alpha \ll \epsilon$, according to Eq. (\ref{Lbom}), at ${\cal
  L}_s < {\cal L}_{b\omega}$ with respect to the fluctuations with
$\omega = \omega_c$, where
\begin{equation}
\label{ps:Lbo} \omega_c = 2 \left( \alpha \over \epsilon \right)^{-2/3}, 
~~~~{\cal L}_{b \omega} = - \frac{\epsilon}{3} \log \alpha \epsilon^2,
\end{equation} 
since in this case $\tilde{l} = \epsilon$.  Similarly, in one dimension
${\cal L}_b = \epsilon \log \epsilon^{-1}$.

As we noted in Sec. IV, there is a one-to-one correspondence between the
AS width ${\cal L}_s$ and the control parameter $A$ . In this system
this correspondence is given by \begin{equation} \label{Ls:pl} A =
{\gamma + e^{-{\cal L}_s \sqrt{1 + \gamma}} \over 2 (1 + \gamma)}
\end{equation} where ${\gamma \over 2( 1 + \gamma)} < A < \frac{1}{2}$.
Thus, knowing the critical values of ${\cal L}_s$, we can easily
calculate the values of $A$ at which the instabilities occur.

Finally, the value of $\beta_\omega$, which appears in Eq. (\ref{beta})
and determines the region where the AS becomes unstable with respect to
pulsations before it transforms into a traveling AS for this model is
\begin{equation} \label{beta:pl} 
  \beta_\omega = 6.2 \cdot 10^{-3} (1 + \gamma)^{-3/2}.
\end{equation} 
This formula also improves the accuracy of the results obtained earlier
in Ref. \cite{kuo:pre95}.

Because of the singular character of the nonlinearity, the potential
defined in Eq. (\ref{V}) is identically zero, so the dispersion relation
for the one-dimensional AS of an arbitrary size in this model becomes
\begin{equation} \label{ps:dis} 
  i \alpha \omega + \epsilon^2 k^2 + \lambda_0 = - {2 \epsilon \over
    \sqrt{1 + \gamma + k^2 + i \omega}} \left\{ 1 \pm e^{ - {\cal L}_s
    \sqrt{1 + \gamma + k^2 + i \omega}} \right\},
\end{equation} 
where the sign ``+'' goes with symmetric fluctuations, whereas the sign
``--'' goes with the antisymmetric ones.  According to Eq.
(\ref{lambda0:calc}),
\begin{equation} \label{ps:lam0} 
  \lambda_0 = - {2 \epsilon \over \sqrt{ 1 + \gamma} } \left\{ 1 - e^{-
    {\cal L}_s \sqrt{1 + \gamma} } \right\}.
\end{equation} 
One can see that Eq. (\ref{ps:dis}) coincides with Eq. (5.4b) of Ref.
\cite{ohta89} by a different method.

As in the case of one-dimensional AS, the unperturbed Green's functions
from Eqs. (\ref{G0:3}) and (\ref{G0:2}) for the spherically- and
cylindrically-symmetric AS, respectively, are the exact Green's
functions. For this reason the exact dispersion relations for the
radially-symmetric AS in this model are given by Eqs. (\ref{d:3}) and
(\ref{disp:2}) with the functions $R_l(\omega)$ and $R_m(k, \omega)$
given by Eqs. (\ref{Rl:3}) and (\ref{R0:2}), respectively. It is easy to
see that the dispersion relations obtained in this fashion coincide with
those obtained in Ref.  \cite{ohta89} as well.

\begin{figure*}[htb]
\begin{center}
\caption{} Qualitative form of the nullclines of Eqs. (\ref{act}) and
(\ref{inh}).
\label{f1}
\end{center}
\end{figure*}

\begin{figure*}[htb]
\begin{center}
\caption{} ``Hot'' and ``cold'' regions forming a pattern. The walls of
the pattern are localized in the region of order $\epsilon$ around
${\cal S}_i$.
\label{domain}
\end{center}
\end{figure*}

\begin{figure*}[htb]
\begin{center}
\caption{} Distributions of $\theta$ and $\eta$ in the form of a
one-dimensional AS. On the AS periphery $\theta$ goes to $\theta_h$ and
$\eta$ goes to $\eta_h$. Dashed lines indicate the regions where the AS
walls are localized.
\label{as}
\end{center}
\end{figure*}

\begin{figure*}[htb]
\begin{center}
\caption{} Stability diagram for the spherically-symmetric AS in the
rescaled variables $\bar{\alpha}$ and $p$. The solid curves correspond
to the thresholds of the instabilities for $l = 0$, $l = 1$, and $l =
2$. The bottom and top horizontal lines correspond to the thresholds of
the aperiodic instabilities for $l = 0$ and $l = 2$, respectively.
  \label{puls3d}
\end{center}
\end{figure*}

\begin{figure*}[htb]
\begin{center}
\caption{} The rescaled frequency $\bar{\omega}$ vs. $\bar{\alpha}$ at
the threshold of the $l = 0$ instability of the spherically-symmetric
AS.
  \label{omegac3d}
\end{center}
\end{figure*}

\begin{figure*}[htb]
\begin{center}
\caption{} Stability diagram for the radially-symmetric AS in two
dimensions in the rescaled variables $\bar{\alpha}$ and $p$. The solid
curves correspond to the thresholds of the instabilities for $m = 0$, $m
= 1$, and $m = 2$. The top horizontal line corresponds to the thresholds
of the aperiodic instability for $l = 2$. The bottom horizontal line
shows schematically the threshold of the aperiodic instability for $m =
0$.
  \label{puls2d}
\end{center}
\end{figure*}

\begin{figure*}[htb]
\begin{center}
\caption{} The rescaled frequency $\bar{\omega}$ vs. $\bar{\alpha}$ at the
threshold of the $m = 0$ instability of the radially-symmetric AS in two
dimensions.
  \label{omegac2d}
\end{center}
\end{figure*}

\end{document}